\documentclass{emulateapj}
\usepackage{colordvi}
\usepackage{epsfig}
\usepackage{graphicx}
\usepackage{amssymb}
\newcommand{\Ha}{\hbox{{\rm H}\kern 0.1em$\alpha$}}
\newcommand{\Hb}{\hbox{{\rm H}\kern 0.1em$\beta$}}
\newcommand{\Hd}{\hbox{{\rm H}\kern 0.1em$\delta$}}
\newcommand{\Hg}{\hbox{{\rm H}\kern 0.1em$\gamma$}}
\newcommand{\Hda}{\hbox{{\rm H}\kern 0.1em$\delta_{\rm{A}}$}}

\newcommand{\MgII}{\hbox{{\rm Mg}\kern 0.1em{\sc ii}}}
\newcommand{\CIV}{\hbox{{\rm C}\kern 0.1em{\sc iv}}}
\newcommand{\NeV}{\hbox{[{\rm Ne}\kern 0.1em{\sc v}]}}
\newcommand{\OII}{\hbox{[{\rm O}\kern 0.1em{\sc ii}]}}
\newcommand{\NeIII}{\hbox{[{\rm Ne}\kern 0.1em{\sc iii}]}}
\newcommand{\OIII}{\hbox{[{\rm O}\kern 0.1em{\sc iii}]}}
\newcommand{\NII}{\hbox{[{\rm N}\kern 0.1em{\sc ii}]}}
\newcommand{\SII}{\hbox{[{\rm S}\kern 0.1em{\sc ii}]}}

\newcommand{\suny}{M$_{\odot}$~yr$^{-1}$}
\newcommand{\lssfr}{log(sSFR/Gyr$^{-1}$)}
\newcommand{\lmass}{log(M/M$_{\odot}$)}

\newcommand{\mdust}{M$_{\rm{dust}}$}

\newcommand{\sigone}{$\Sigma_{1}$}

\begin{document}

\title{Sub-kpc ALMA imaging of compact star-forming galaxies at
  $\lowercase{z}\sim2.5$: revealing the formation of dense galactic cores in the progenitors of compact quiescent galaxies}
\author{G. Barro\altaffilmark{1},
M. Kriek\altaffilmark{1},
P. G.~P\'{e}rez-Gonz\'{a}lez\altaffilmark{2},
J. R.~Trump\altaffilmark{3,4},
D. C.~Koo\altaffilmark{5}, 
S. M.~Faber\altaffilmark{5},
A. Dekel\altaffilmark{6},
J. R.~Primack\altaffilmark{7},
Y. Guo\altaffilmark{3},
D. D.~Kocevski\altaffilmark{8}, 
J. C.~Mu{\~n}oz-Mateos\altaffilmark{9},
W. Rujoparkarn\altaffilmark{10},
K. Sheth\altaffilmark{11}}

\altaffiltext{1}{University of California Berkeley}
\altaffiltext{2}{Universidad Complutense de Madrid}
\altaffiltext{3}{Pennsylvania State University}
\altaffiltext{4}{Hubble Fellow}
\altaffiltext{5}{University of California Santa Cruz}
\altaffiltext{6}{The Hebrew University}
\altaffiltext{7}{Santa Cruz Institute for Particle Physics}
\altaffiltext{8}{Colby College}
\altaffiltext{9}{European Southern Observatory}
\altaffiltext{10}{Kavli Institute for the Physics of the Universe}
\altaffiltext{11}{National Radio Astronomy Observatory}

\slugcomment{To be submitted to the Astrophysical Journal} 
\slugcomment{Last edited: \today}
\label{firstpage}
\begin{abstract} 

  { We present spatially-resolved Atacama Large
    Millimeter/sub-millimeter Array (ALMA) 870~$\mu$m dust continuum
    maps of six massive, compact, dusty star-forming galaxies (SFGs)
    at $z\sim2.5$.  These galaxies are selected for their small
    rest-frame optical sizes ($r_{\rm e, F160W}\sim1.6$~kpc) and high
    stellar-mass densities that suggest that they are direct
    progenitors of compact quiescent galaxies at $z\sim2$. The deep
    observations yield high far-infrared (FIR) luminosities of L$_{\rm
      IR}=10^{12.3-12.8}$~L$_{\odot}$ and star formation rates (SFRs)
    of SFR~$=200-700$~\suny, consistent with those of typical
    star-forming ``main sequence'' galaxies. The high-spatial
    resolution (FWHM$\sim0\farcs12-0\farcs18$) ALMA and HST photometry
    are combined to construct deconvolved, mean radial profiles of
    their stellar mass and (UV+IR) SFR. We find that the dusty,
    nuclear IR-SFR overwhelmingly dominates the bolometric SFR up to
    $r\sim5$~kpc, by a factor of over 100$\times$ from the unobscured
    UV-SFR. Furthermore, the effective radius of the mean SFR profile
    ($r_{\rm e, SFR}\sim1$~kpc) is $\sim$30\% smaller than that of the
    stellar mass profile. The implied structural evolution, if such
    nuclear starburst last for the estimated gas depletion time of
    $\Delta t=\pm100$~Myr, is a 4$\times$ increase of the stellar mass
    density within the central 1~kpc and a 1.6$\times$ decrease of the
    half-mass radius. This structural evolution fully supports
    dissipation-driven, formation scenarios in which strong nuclear
    starbursts transform larger, star-forming progenitors into compact
    quiescent galaxies.}

\end{abstract}

\section{Introduction}\label{intro}

\begin{figure*}[t]
  \centering
  \includegraphics[width=8.1cm,angle=0.]{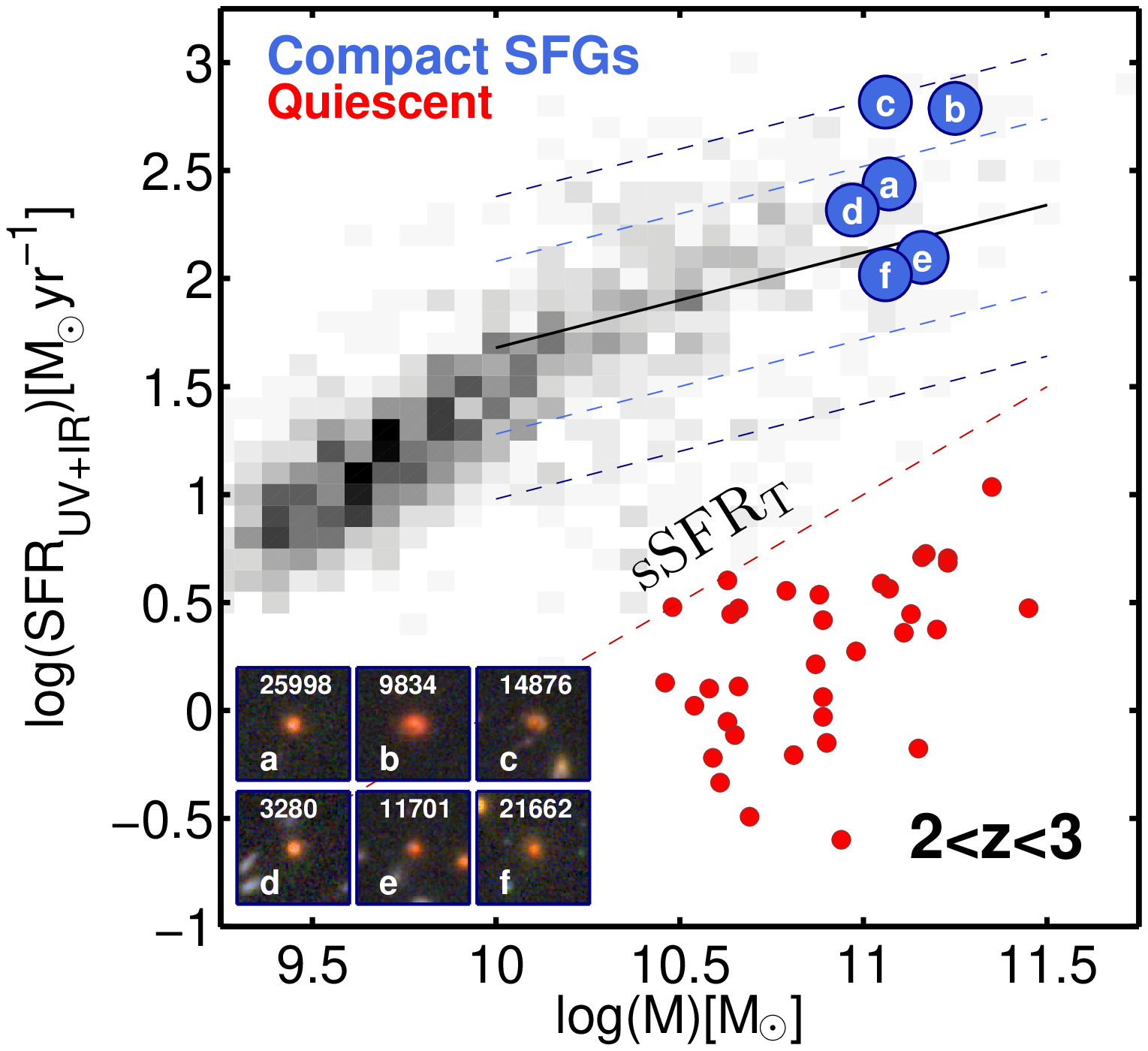}
  \hspace{1cm}
  \includegraphics[width=8cm,angle=0.]{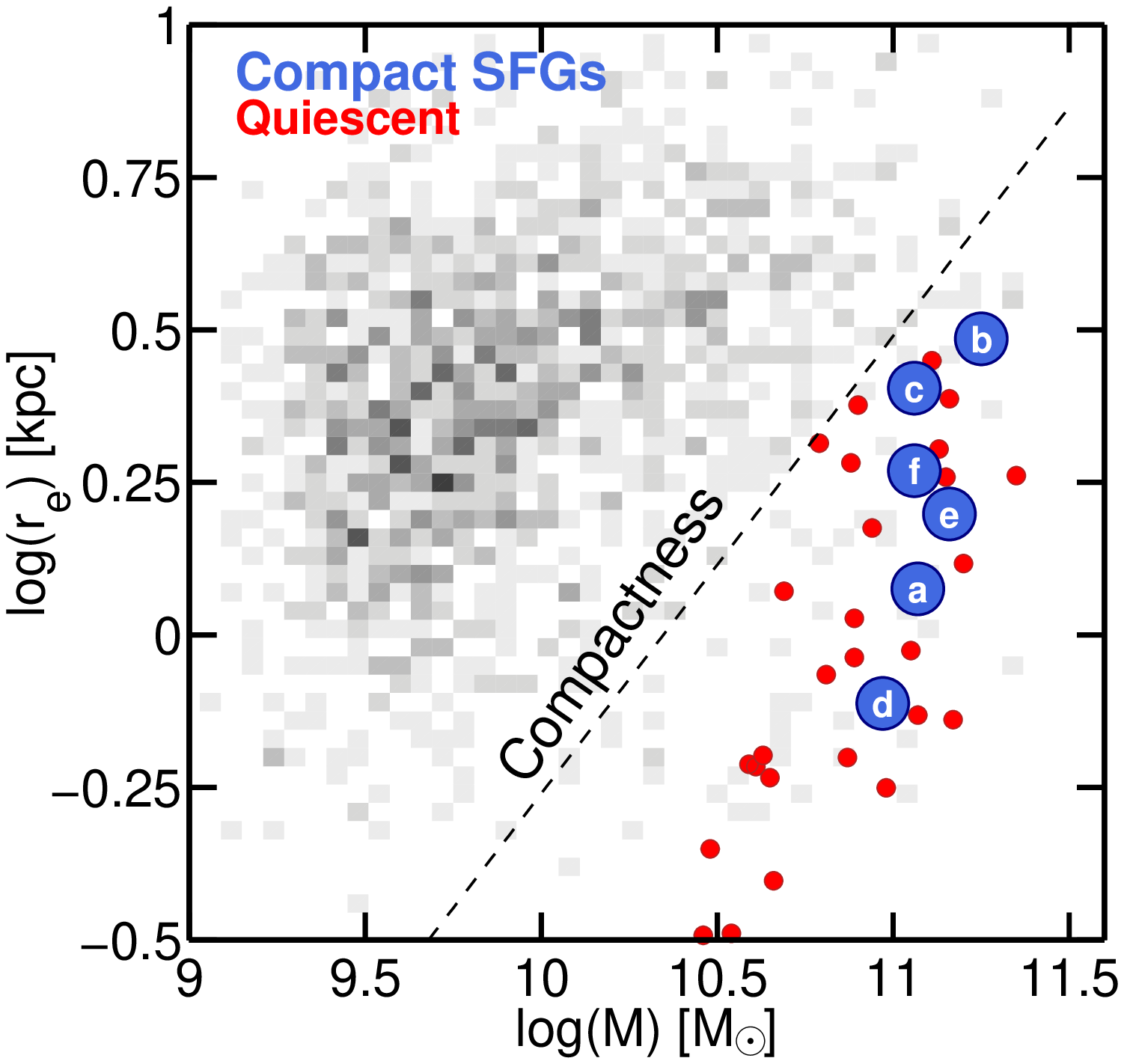}
\caption{\label{selectiondiag} {\it Left:} SFR--mass diagram for
  galaxies in CANDELS GOODS-S at $2<z<3$.  The grey-scale density bins
  map the location of the SFR-MS. The solid black and dashed blue
  lines depict the best fit and 2.5$\times$ and 5$\times$ limits above
  and below the SFR-MS. The blue circles depict the compact SFGs
  observed with ALMA. The subpanels in the bottom-left corner show the
  $5''\times5''$ ACS/WFC3 $zJH$ images of the ALMA galaxies. The red
  dashed line marks the threshold in sSFR (\lssfr$<-1$) used to
  identify quiescent galaxies (red circles). {\it Right:} mass--size
  distribution for the same galaxies as in the left panel. The dashed
  line marks the {\it compactness} threshold,
  log$(\Sigma_{1.5})=10.4~M_{\odot}$kpc$^{-1.5}$.}
\end{figure*}

The majority of SFGs follow a relatively tight, almost linear relation
between SFR and stellar mass, usually referred to as the
star-formation ``main sequence'' that seems to be in place since
$z\sim5-6$ (SF-MS; e.g., \citealt{mainseq};
\citealt{whitaker12b}). The ubiquitous and tight SF-MS suggests that
the majority of the stars are formed in a predominantly smooth, {\it
  secular} mode.  Furthermore, there is also evidence that, despite
their wide range of sizes and morphologies, most of the stars in SFGs
are formed in disks which are growing from the inside out, thus
increasing their sizes with cosmic time (\citealt{wuyts13};
\citealt{nelson13, nelson15}).  The progressive structural growth in
the SF-MS is consistent with the classic notion of galaxy formation in
a $\Lambda$CDM Universe in which gas accreted from dark matter halos
cools and forms new stars in disks with increasingly larger scale
lengths with cosmic time (e.g., \citealt{fall80}; \citealt{mo98}).

A challenge to this simplified picture are the small sizes
($r_{e}\sim1$~kpc) of the first massive quiescent galaxies at
$z\gtrsim1.5-3$ (e.g., \citealt{vdw14} and references therein).  On
one hand, their small sizes might be the consequence of having smaller
star-forming progenitors formed at earlier times when the Universe was
more dense (i.e., more concentrated haloes and higher gas
fractions). On the other hand, compact quiescent galaxies could form
in strongly dissipative processes, triggered by mergers or
interaction-driven disk instabilities that cause a substantial growth
of the nuclear stellar density as a result of gas-rich starbursts
(\citealt{hopkins08a}; \citealt{dekel09b}). Both scenarios imply the
formation of compact SFGs as the last stage before quenching star
formation, but the predictions differ on whether these compact SFGs
would exhibit extended SFR profiles, driving the inside-out size
growth, or compact star-forming regions (starbursts) triggered
by the dissipative phase.

Such compact SFGs have been identified in sizable numbers and their
small stellar sizes, steep mass profiles and obscured SFR properties
have been confirmed by multiple studies (e.g., \citealt{barro13,
  barro14a}; \citealt{dokkum15}). However, direct measurements of
their spatial distribution of the star formation relative to the mass
profile, needed to discriminate between the two formation scenarios
discussed above, are still inconclusive. These measurements have
proven very difficult because even spatially resolved UV and optical
SFR indicators based on HST observations are significantly affected by
the high dust obscuration, particularly in galaxy centers
(\citealt{wuyts12}; \citealt{tacchella15}), and FIR observations,
sensitive to ionizing radiation re-emitted by the dust, usually have
very poor spatial resolution. Modern (sub-)millimeter/radio
interferometers such as ALMA and JVLA have opened a new window into
this regime and enable us to measure the dust emission with high
sensitivity and similar spatial resolution as those from HST
observations.

Here, we exploit a joint analysis of the high spatial resolution
HST/ACS and WFC3 and ALMA continuum imaging to simultaneously
characterize the UV- and IR- SFR profiles and the stellar mass
profiles of 6 compact SFGs at $z\sim2.5$. Throughout this paper, we
quote magnitudes in the AB system, assume a \cite{chabrier} initial
mass function (IMF), and adopt the following cosmological parameters:
($\Omega_{M}$,$\Omega_{\Lambda}$,$h$) = (0.3, 0.7, 0.7).

\section{Data and Sample selection}

\begin{figure*}[t]
\includegraphics[width=6cm,angle=0.]{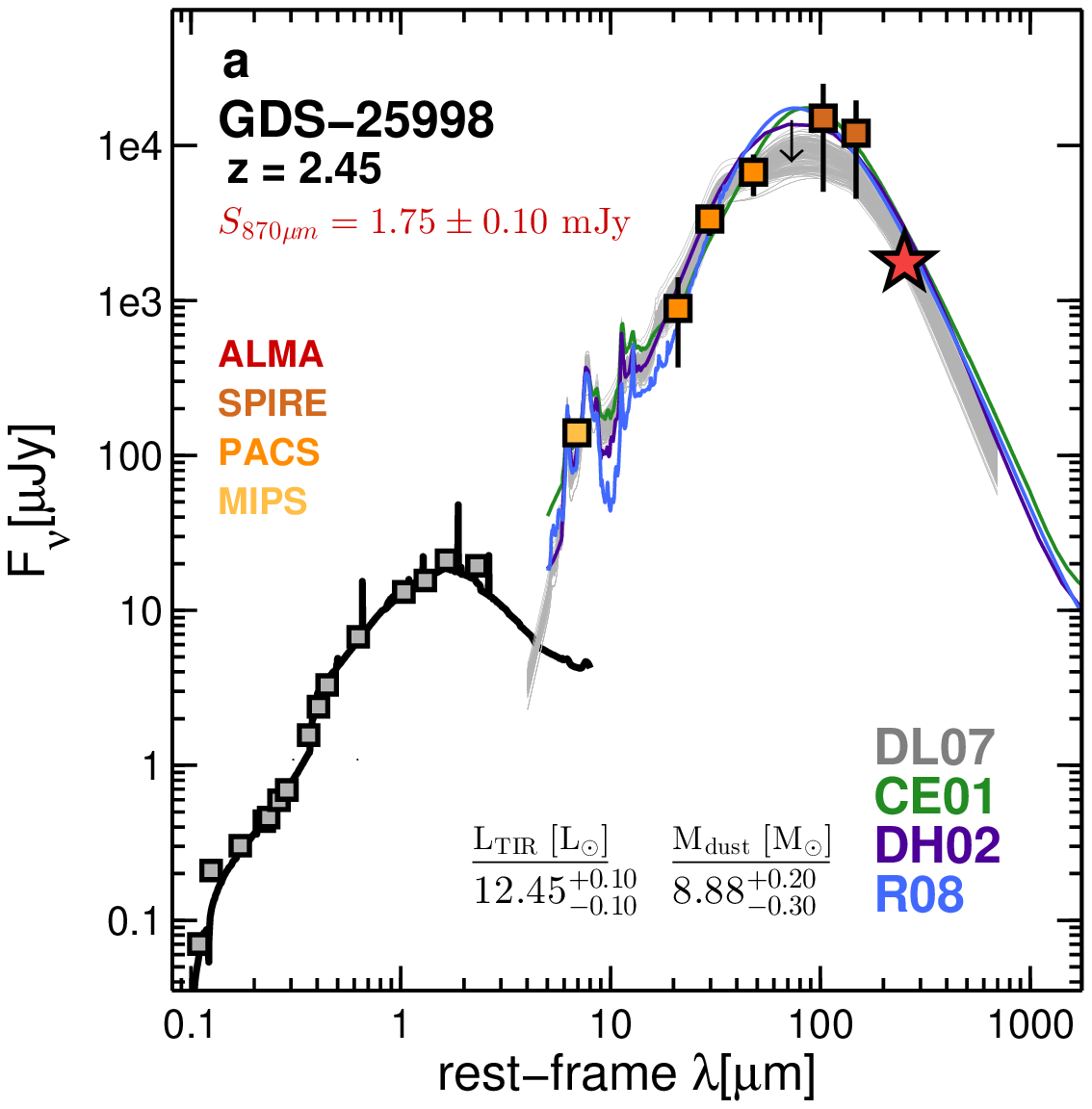}
\includegraphics[width=6cm,angle=0.]{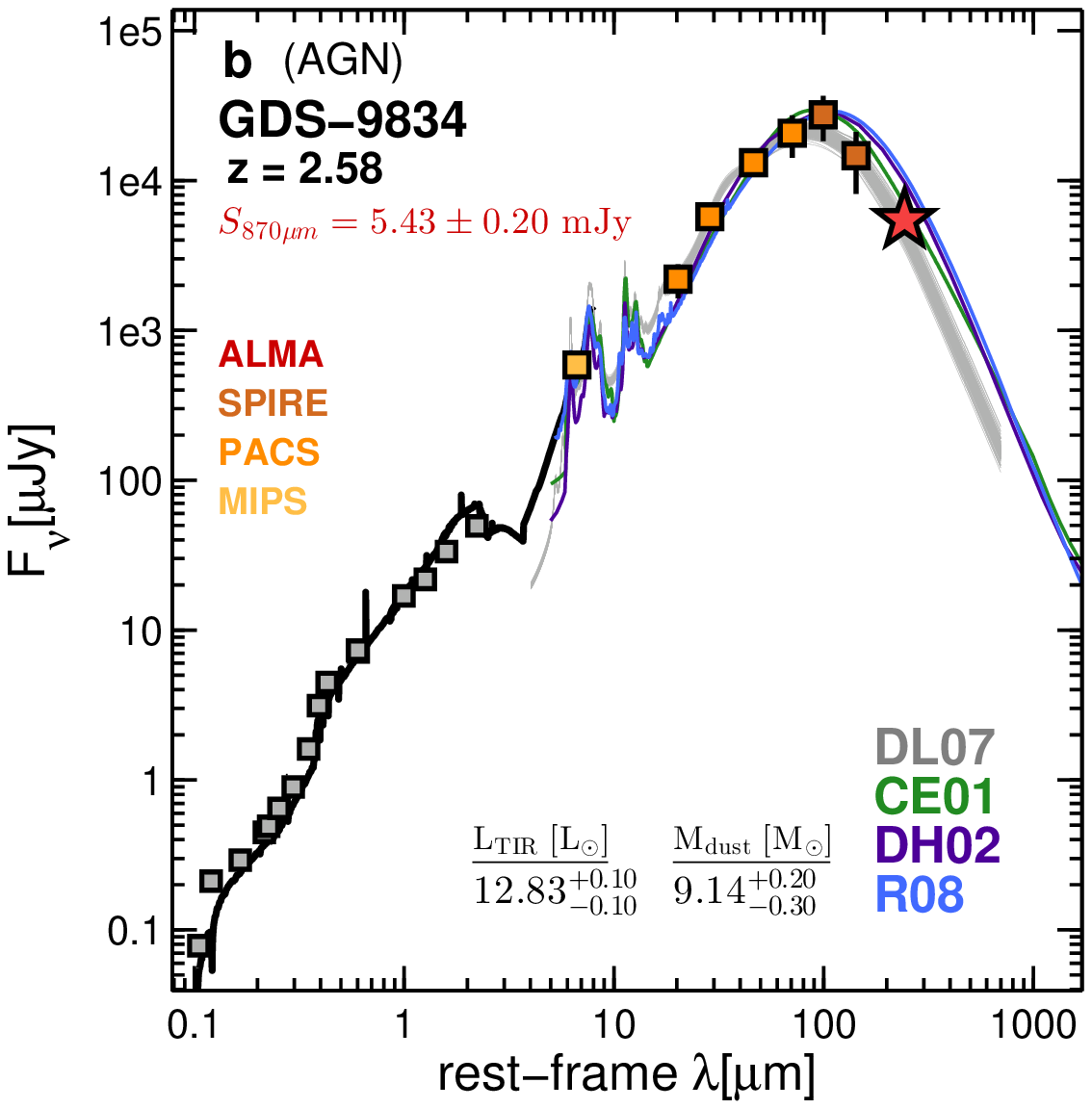}
\includegraphics[width=6cm,angle=0.]{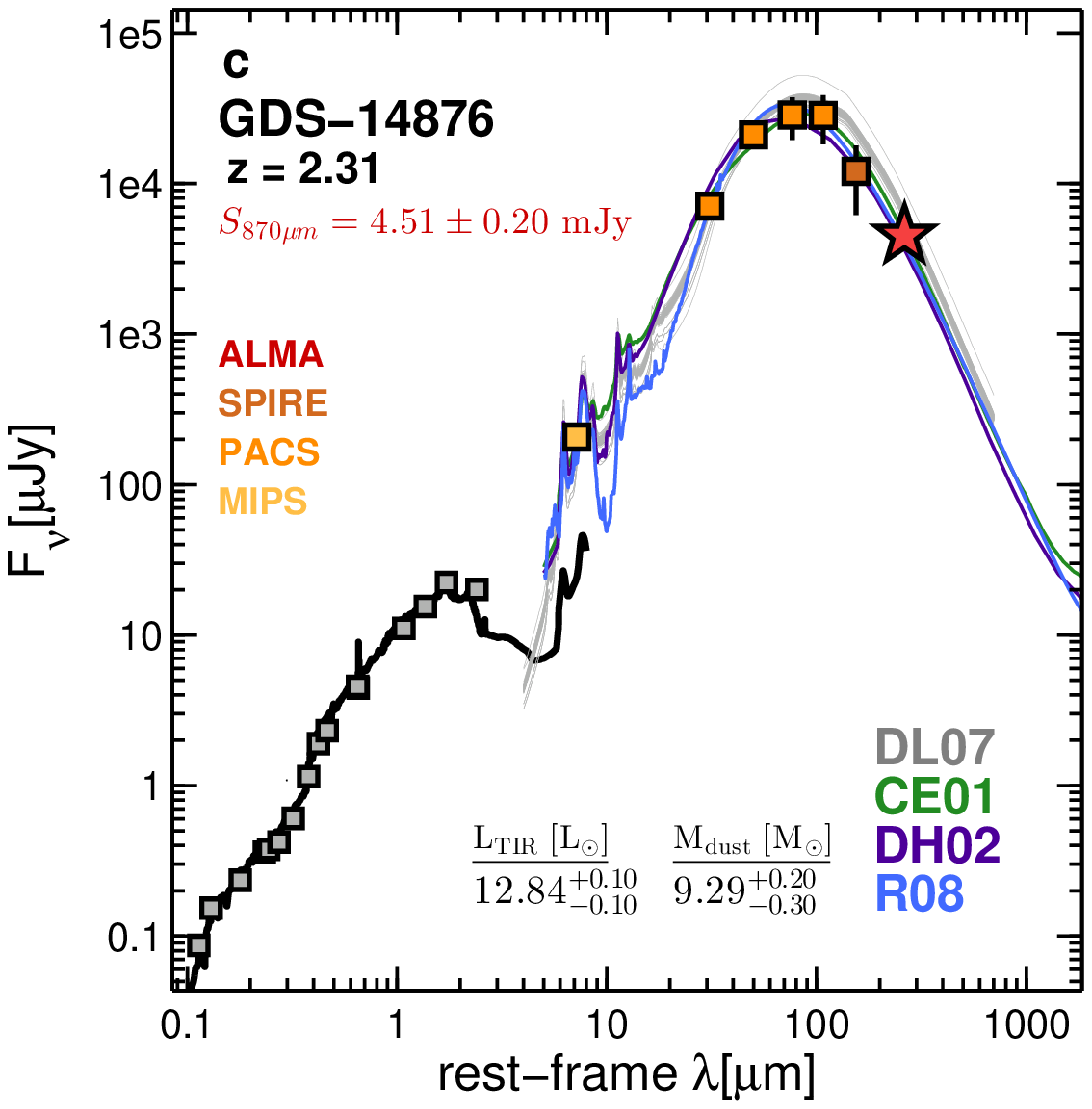}\\
\includegraphics[width=6cm,angle=0.]{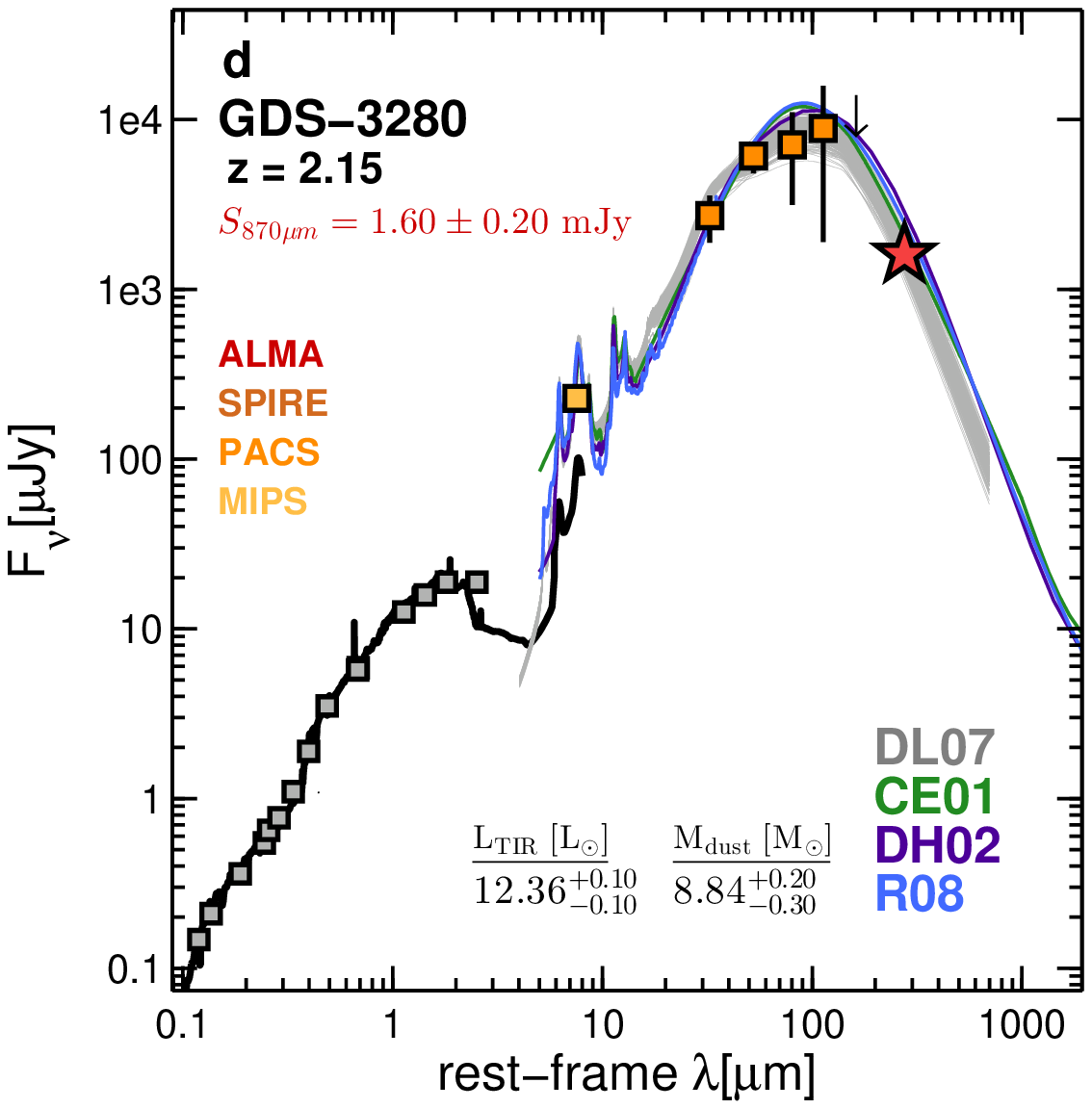}
\includegraphics[width=6cm,angle=0.]{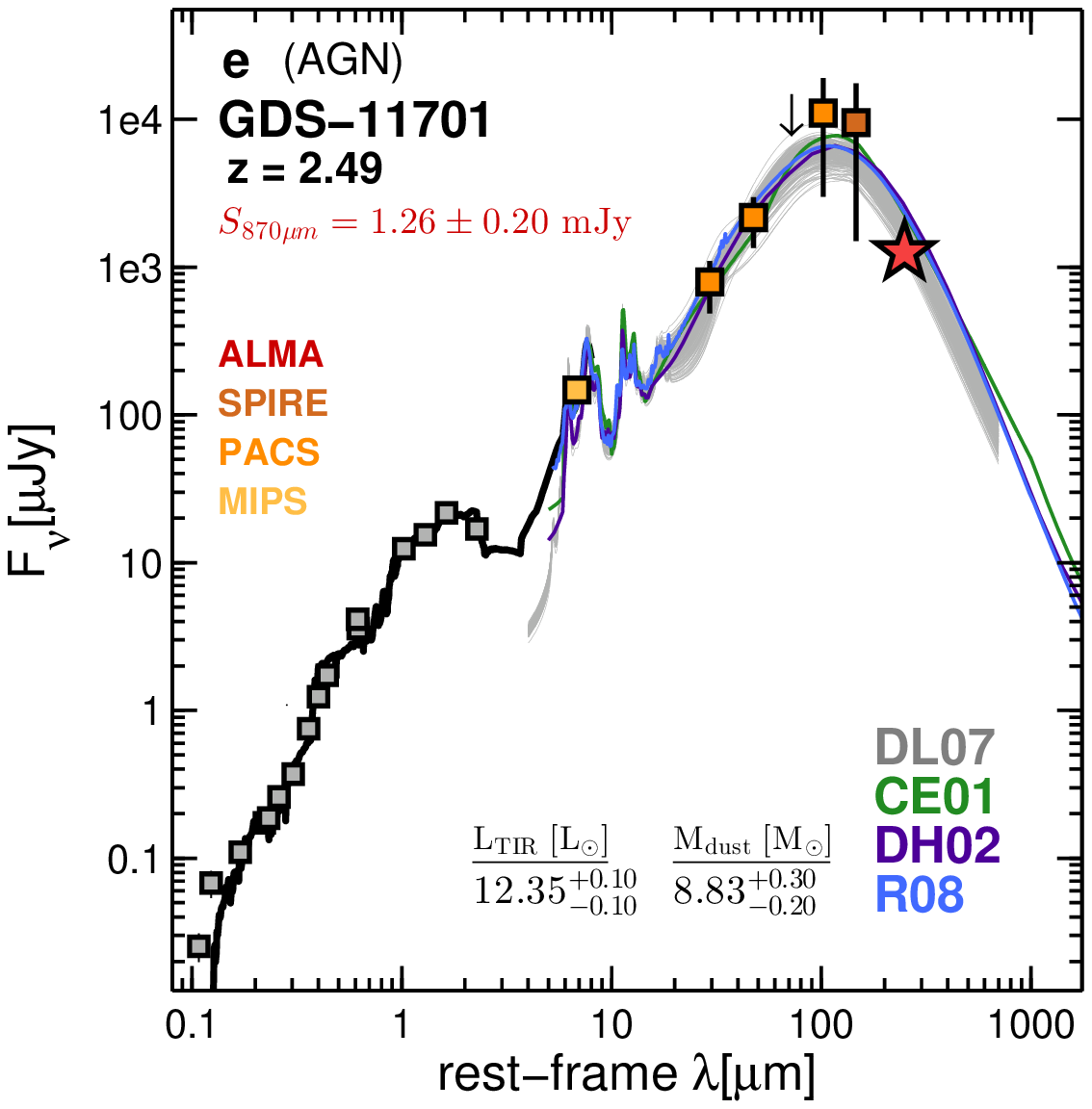}
\includegraphics[width=6cm,angle=0.]{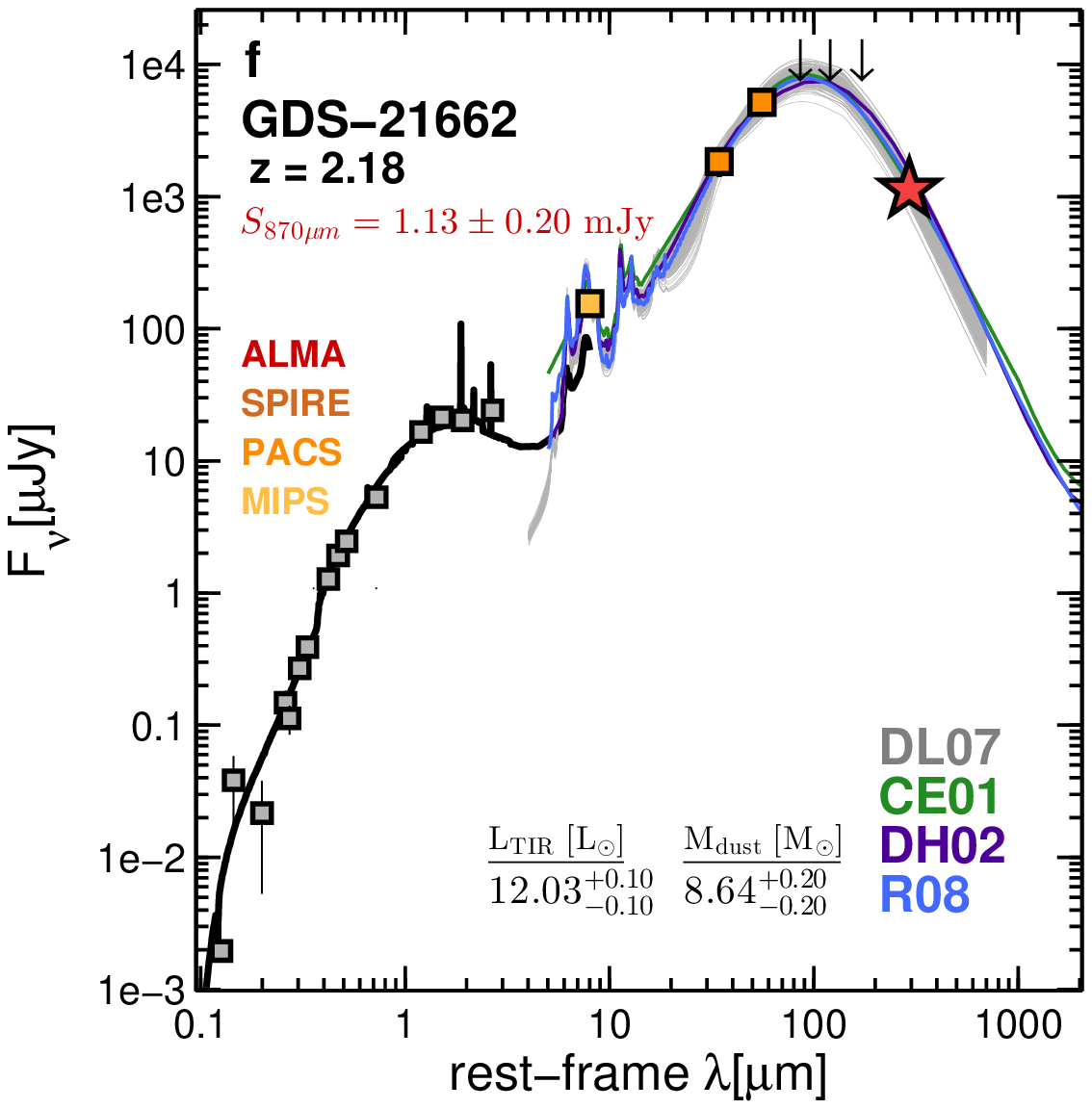}
\caption{\label{seds} UV-to-FIR SEDs of the compact SFGs. The black
  lines show the best-fit BC03 stellar population models for the
  photometry up to 8$\mu$m rest-frame (gray squares), which provide an
  estimate of the stellar population properties and the dust
  attenuation. The orange squares show the mid-to-far IR data from
  {\it Spitzer} MIPS and {\it Herschel} PACS and SPIRE; the red star
  shows the ALMA~870~$\mu$m flux. The green, purple and blue lines
  show the best-fit dust emission models from the libraries of
  \citet{ce01}, \citet{dh02}, \citet{rieke09}. The grey regions depict
  300 models drawn from the posterior probability distribution of the
  {\tt emcee} fit to the \citet{dl07} models. The median values and
  confidence intervals for L$_{\rm TIR}$ and M$_{\rm dust}$ are
  indicated.}
\end{figure*} 

The 6 galaxies analyzed in this paper are drawn from the sample of
compact SFGs in the CANDELS (\citealt{candelsgro}) GOODS-S region
presented in \citet{barro14a}. The UV to near-IR spectral energy
distributions (SEDs) include extensive multi-band data ranging from U
to 8$\mu$m \citep{guo13}.  Furthermore, we include {\it Spitzer}/MIPS
24 and 70~$\mu$m data (30~$\mu$Jy and 1~mJy, $5\sigma$) from
\citet{pg08b}, and PACS 70, 100 and 160~$\mu$m ($0.7$~mJy, $5\sigma$),
and SPIRE 250, 350 and 500~$\mu$m (1~mJy, $5\sigma$) from the
GOODS-Herschel (\citealt{elbaz11}) and PEP (\citealt{magnelli13})
surveys.

The compact SFGs were identified following the method described in
\citet{barro13,barro14a}. Briefly, we require galaxies to be massive
(\lmass$>10.5$), star-forming (\lssfr$>-1$), and we impose a {\it
  compactness} criterion \citep{barro13}, log$(M/\pi
r^{1.5}_{\mathrm{e}})=10.4~M_{\odot}$kpc$^{-1.5}$, to identify
galaxies with similar structural properties as quiescent galaxies at
that redshift. Lastly, we choose FIR bright galaxies detected in {\it
  Spitzer} and {\it Herschel} with predicted ALMA~870~$\mu$m fluxes
above $\sim$1~mJy.  Figure~\ref{selectiondiag} illustrates the
selection criteria by showing the location of the compact SFGs (blue
circles) observed with ALMA overlaid in the SFR-mass and mass-size
diagrams for galaxies more massive than \lmass$>9$ at $2<z<3$ in the
CANDELS GOODS-S catalog.

\begin{figure*}[t]
\centering
\includegraphics[width=4.2cm,angle=0.]{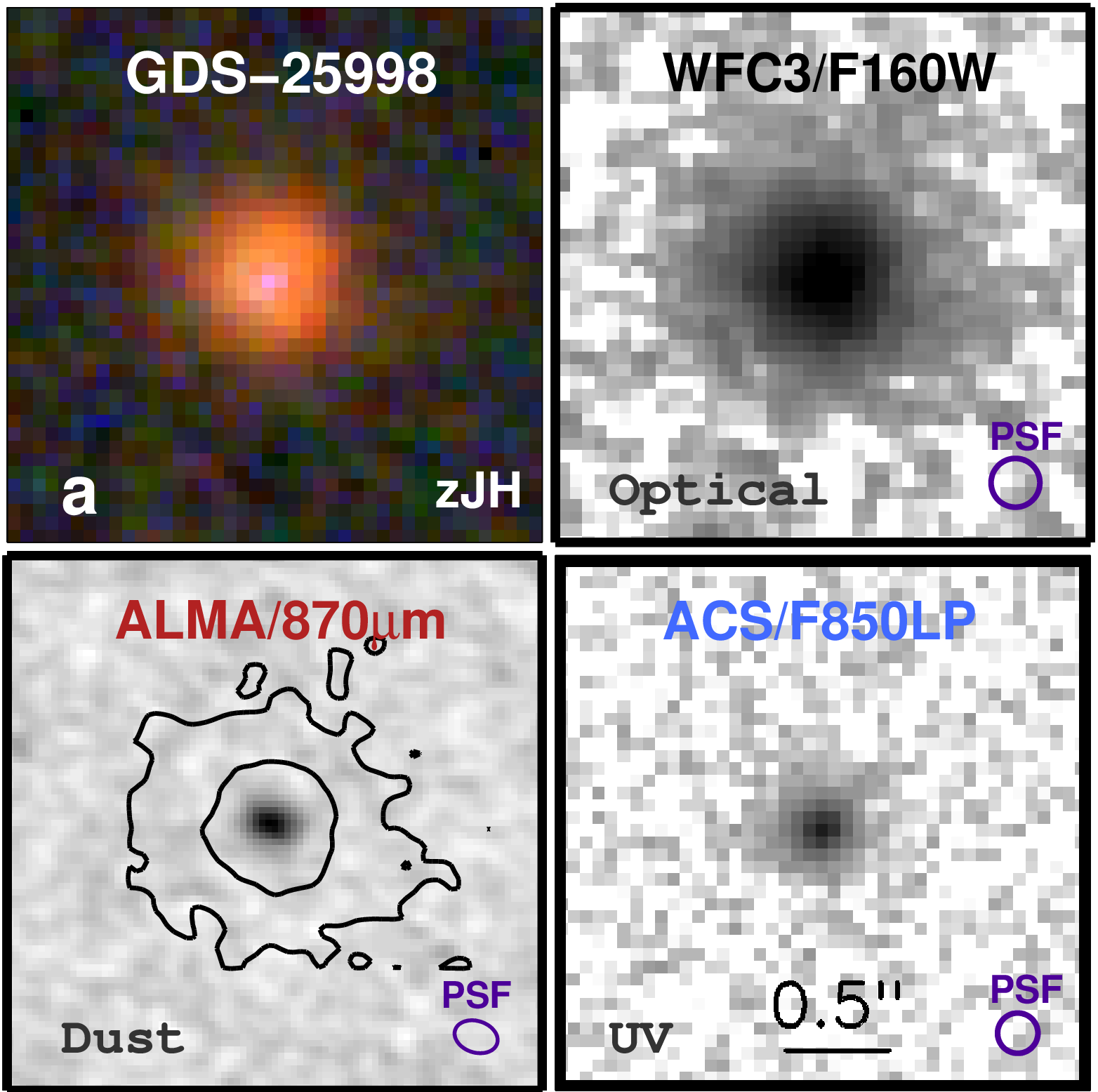}
\includegraphics[width=4.6cm,angle=0.]{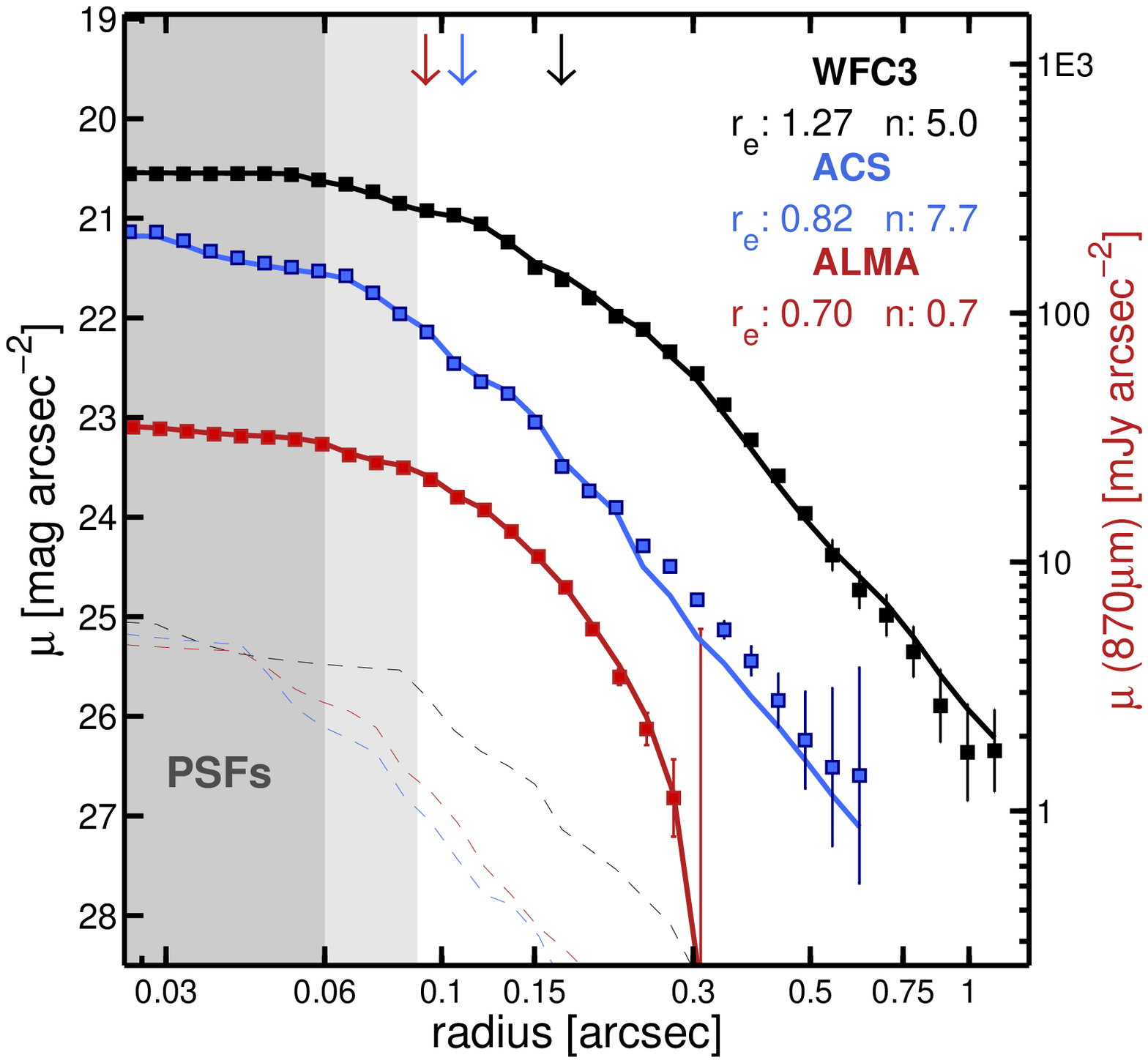}
\hspace{0.cm}
\includegraphics[width=4.2cm,angle=0.]{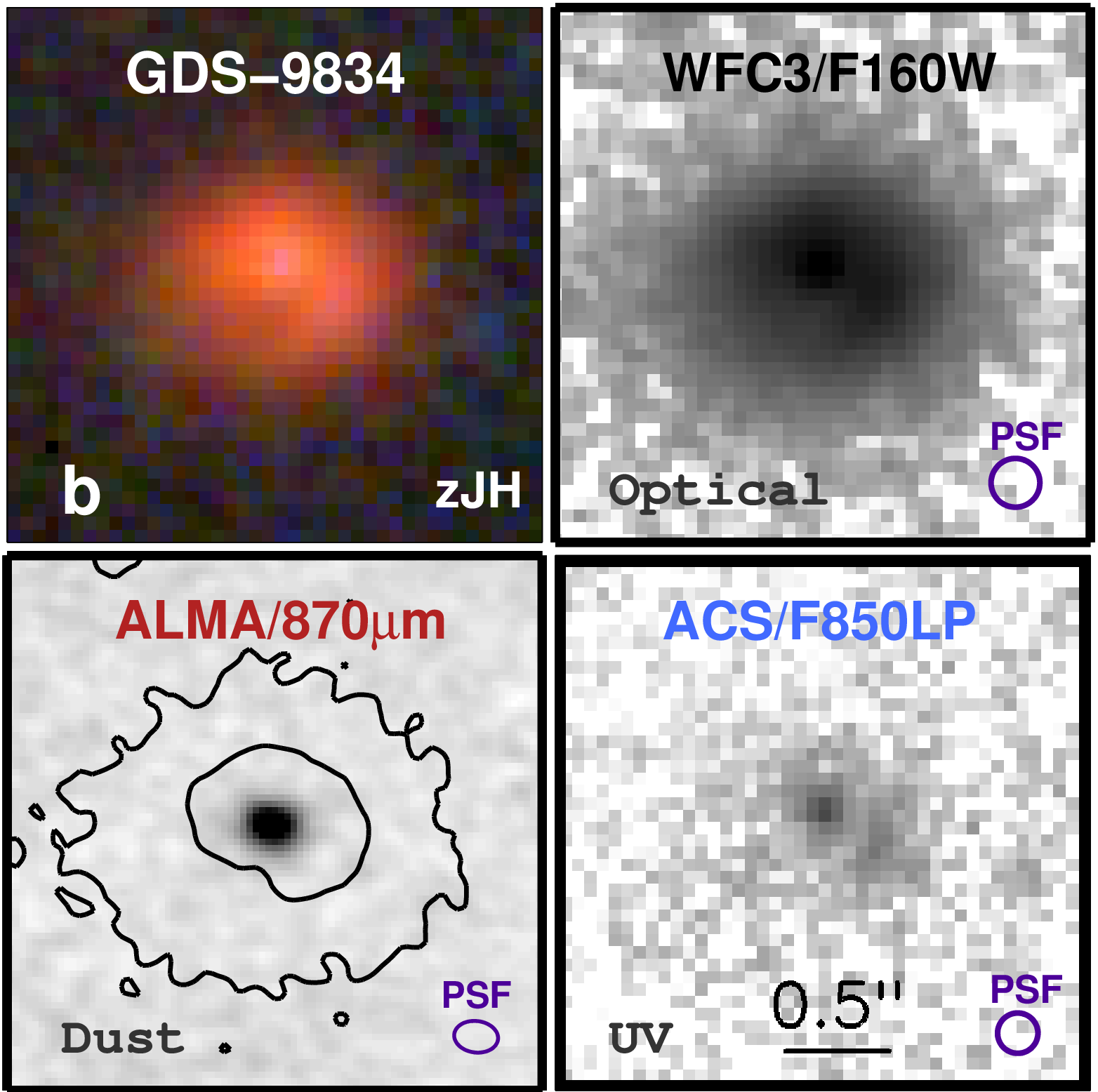}
\includegraphics[width=4.6cm,angle=0.]{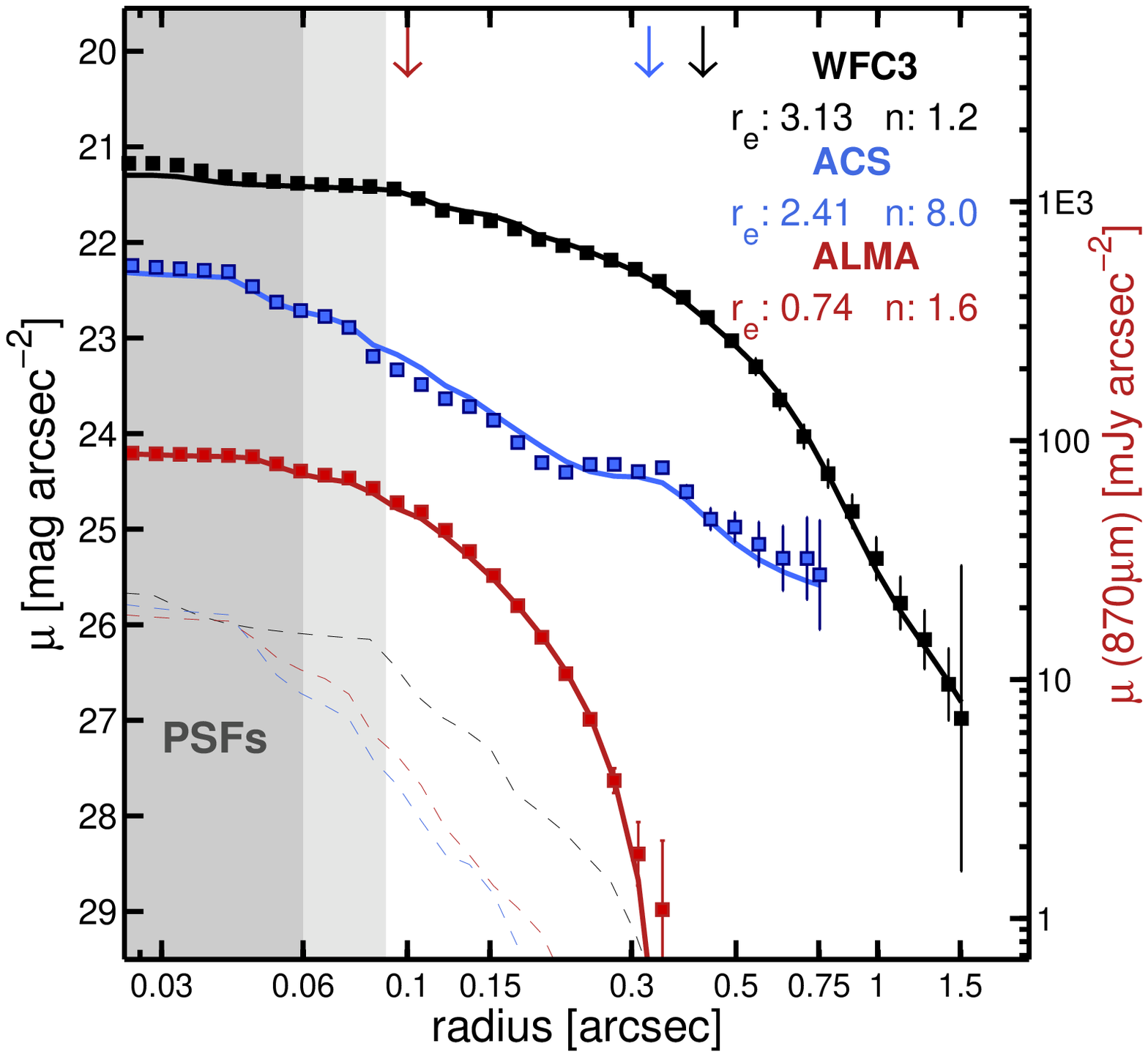}
\includegraphics[width=4.2cm,angle=0.]{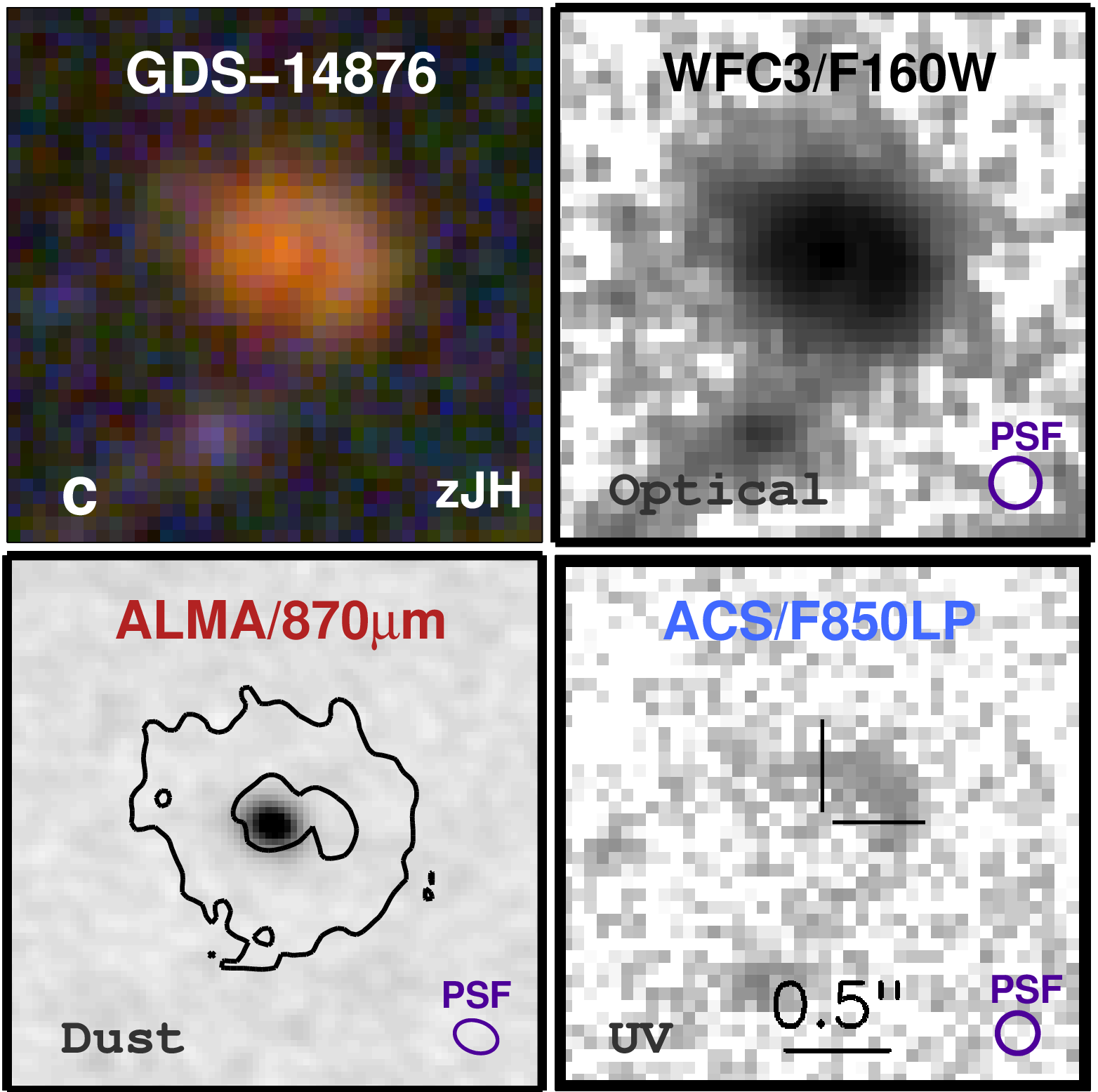}
\includegraphics[width=4.6cm,angle=0.]{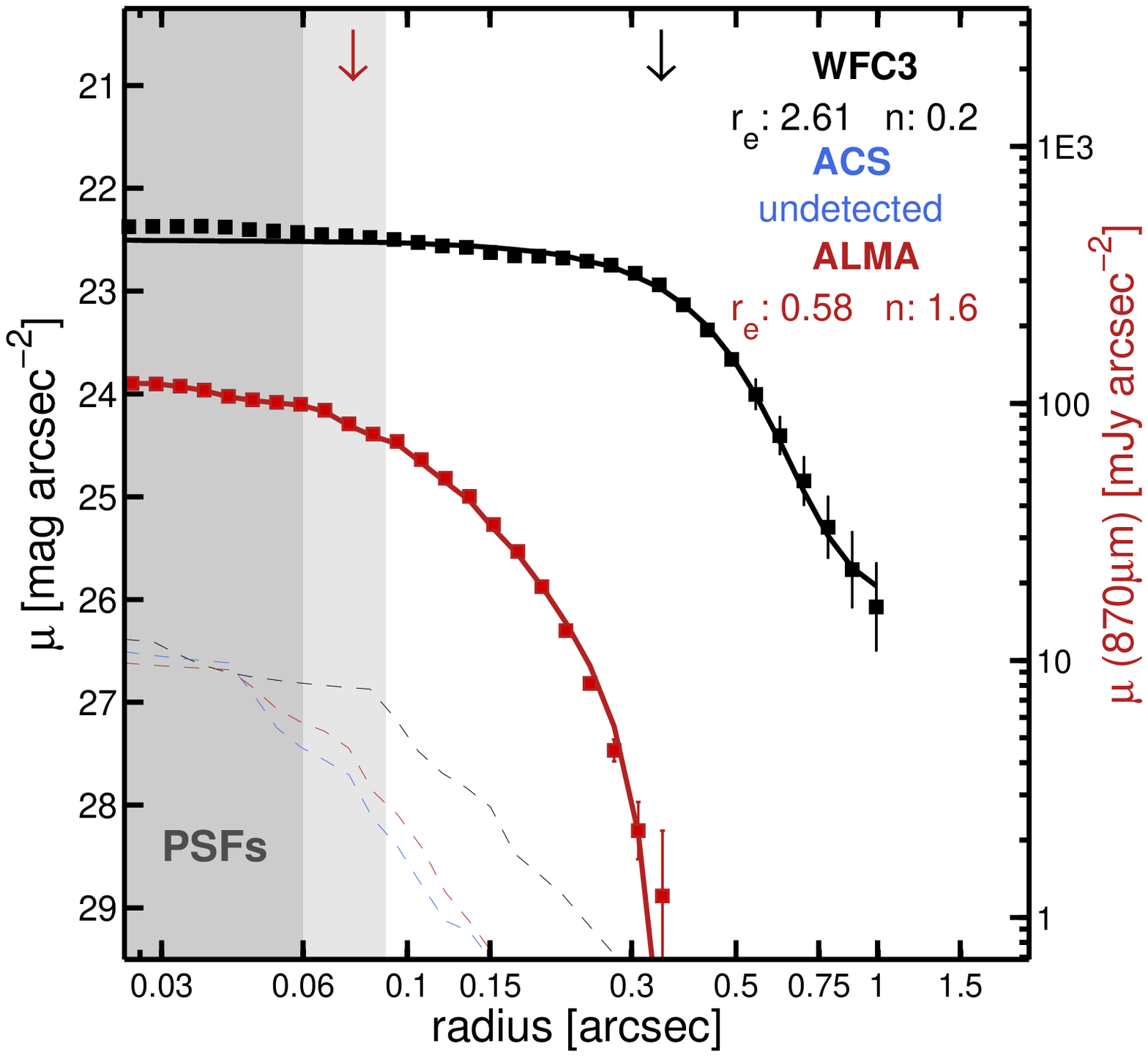}
\hspace{0.cm}
\includegraphics[width=4.2cm,angle=0.]{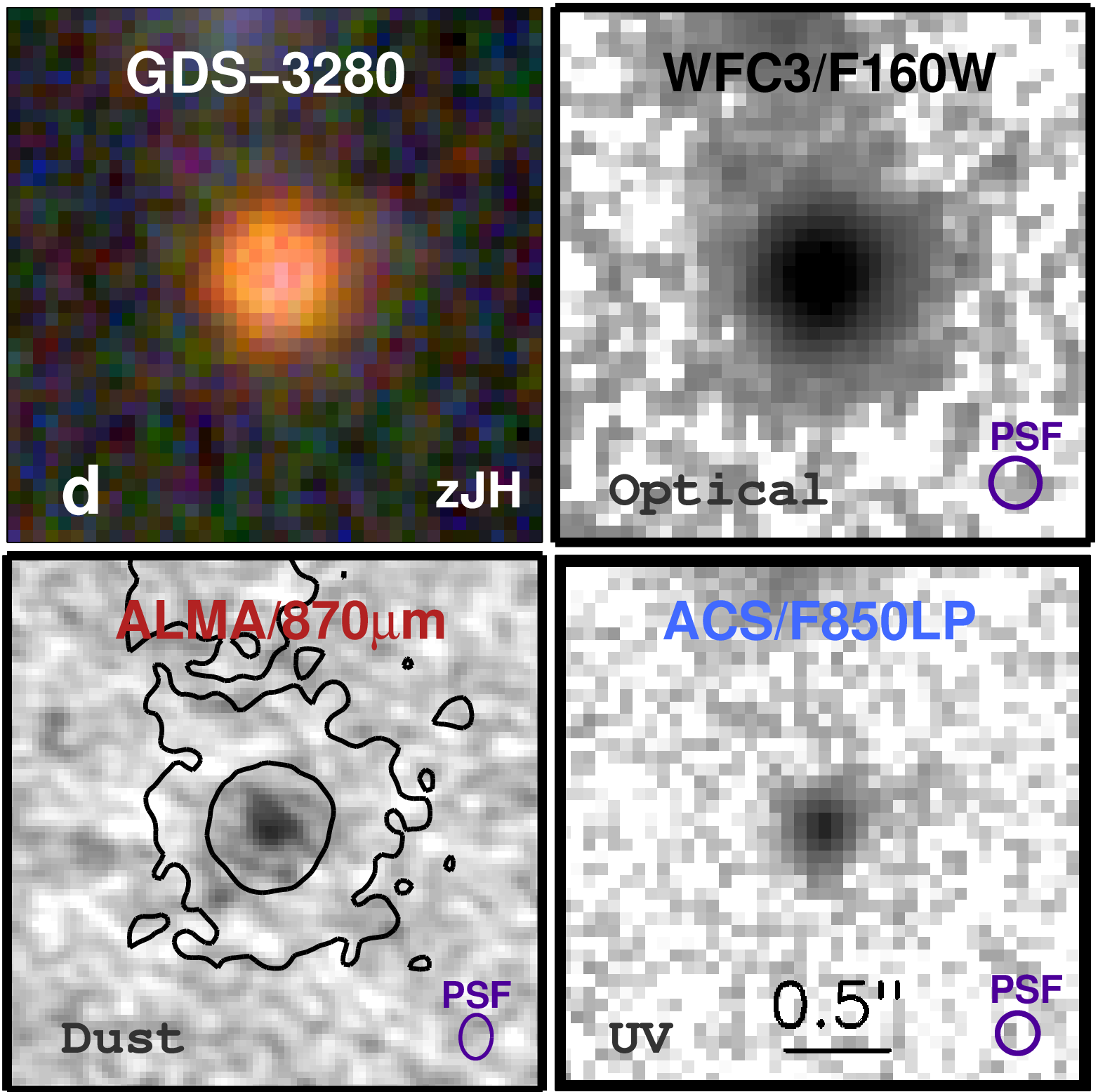}
\includegraphics[width=4.6cm,angle=0.]{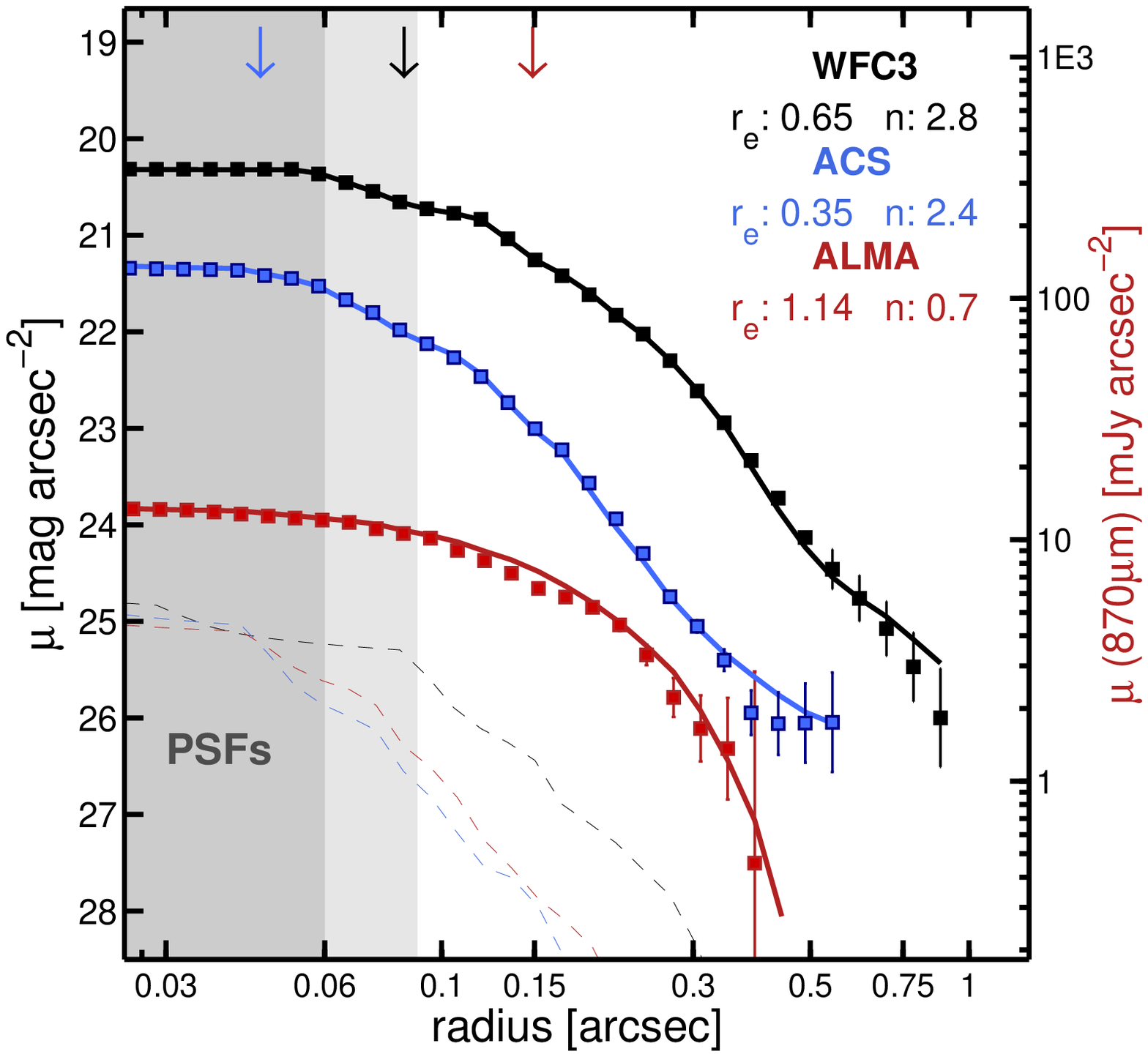}
\includegraphics[width=4.2cm,angle=0.]{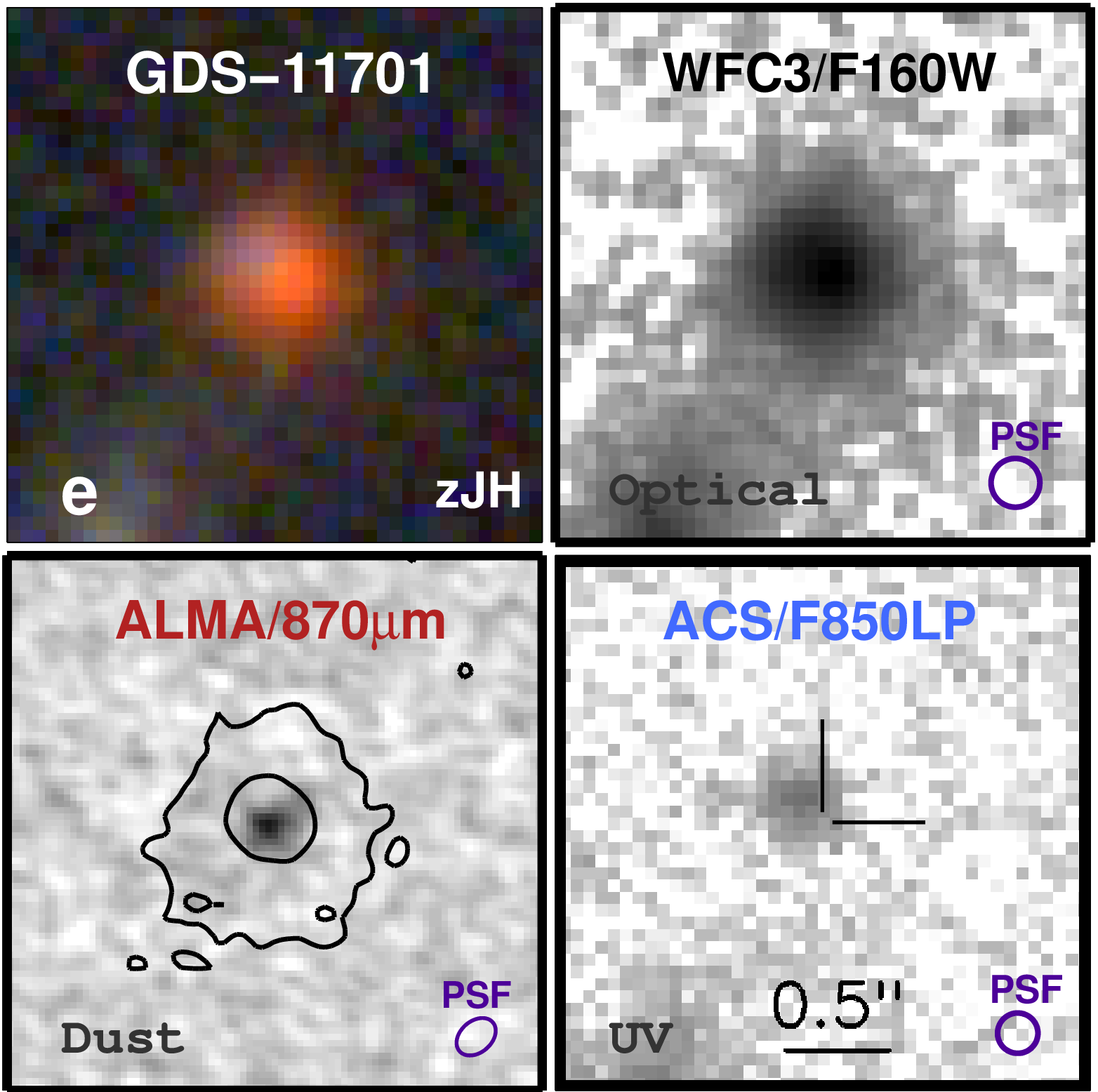}
\includegraphics[width=4.6cm,angle=0.]{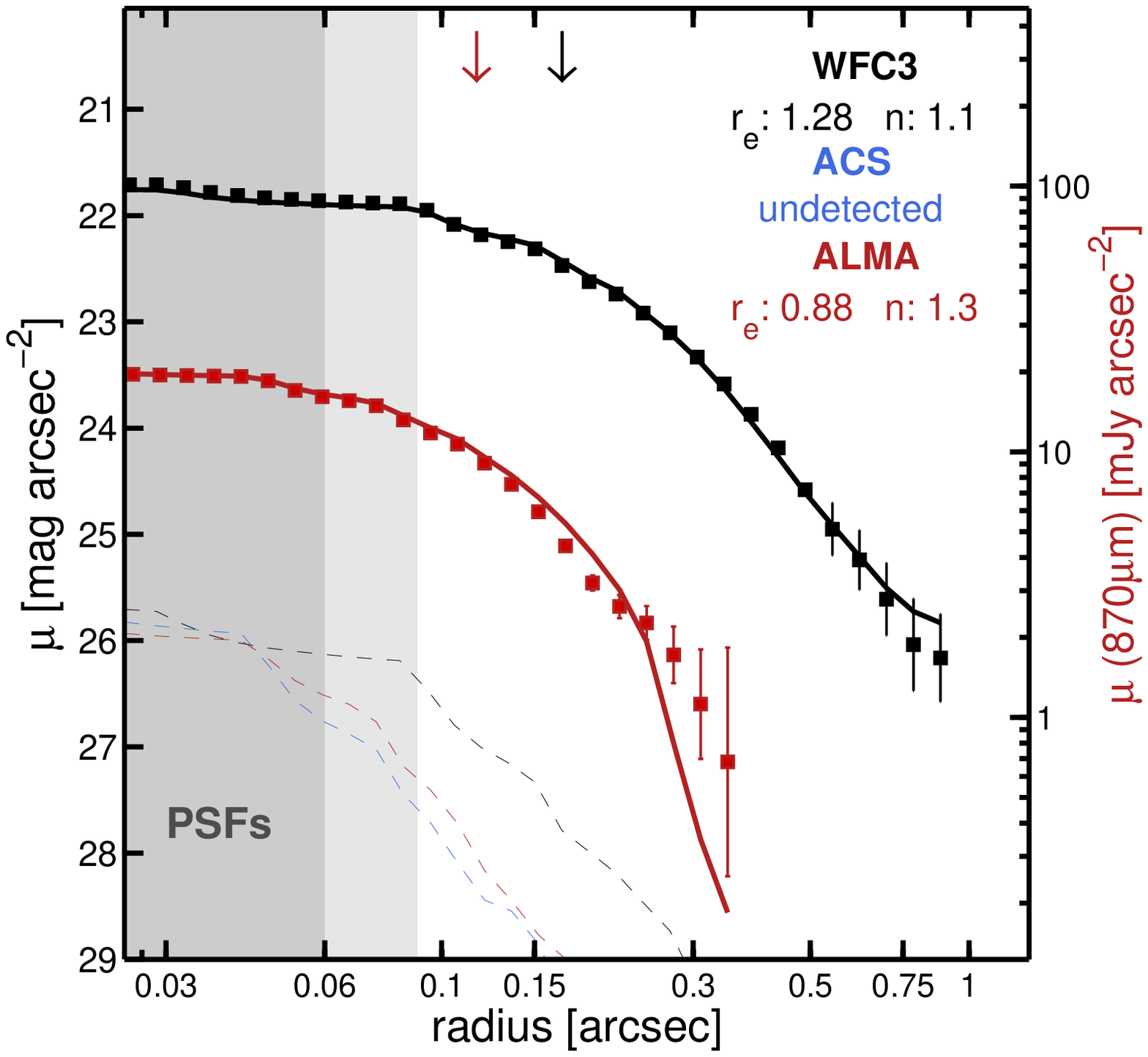}
\hspace{0.cm}
\includegraphics[width=4.2cm,angle=0.]{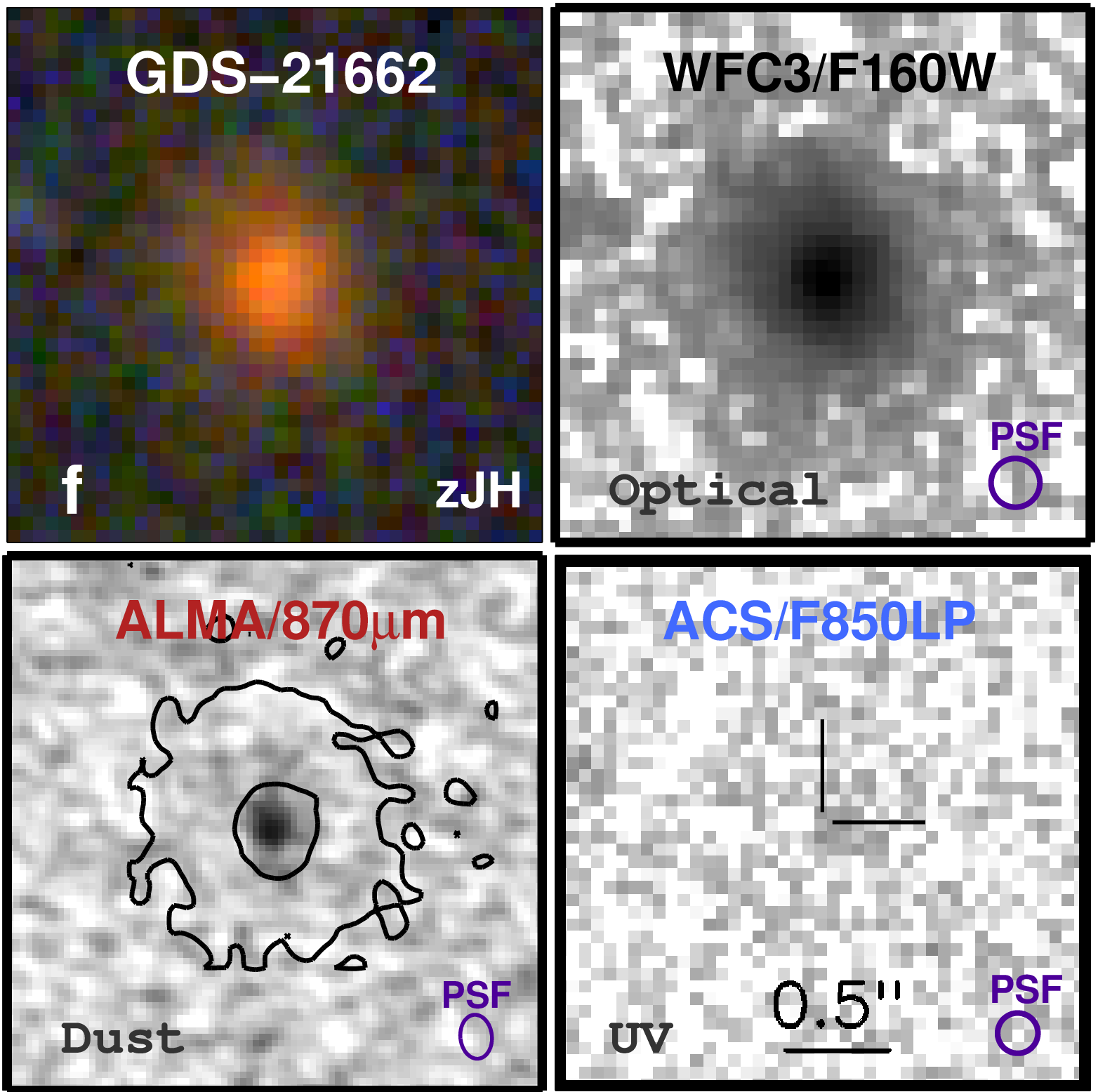}
\includegraphics[width=4.6cm,angle=0.]{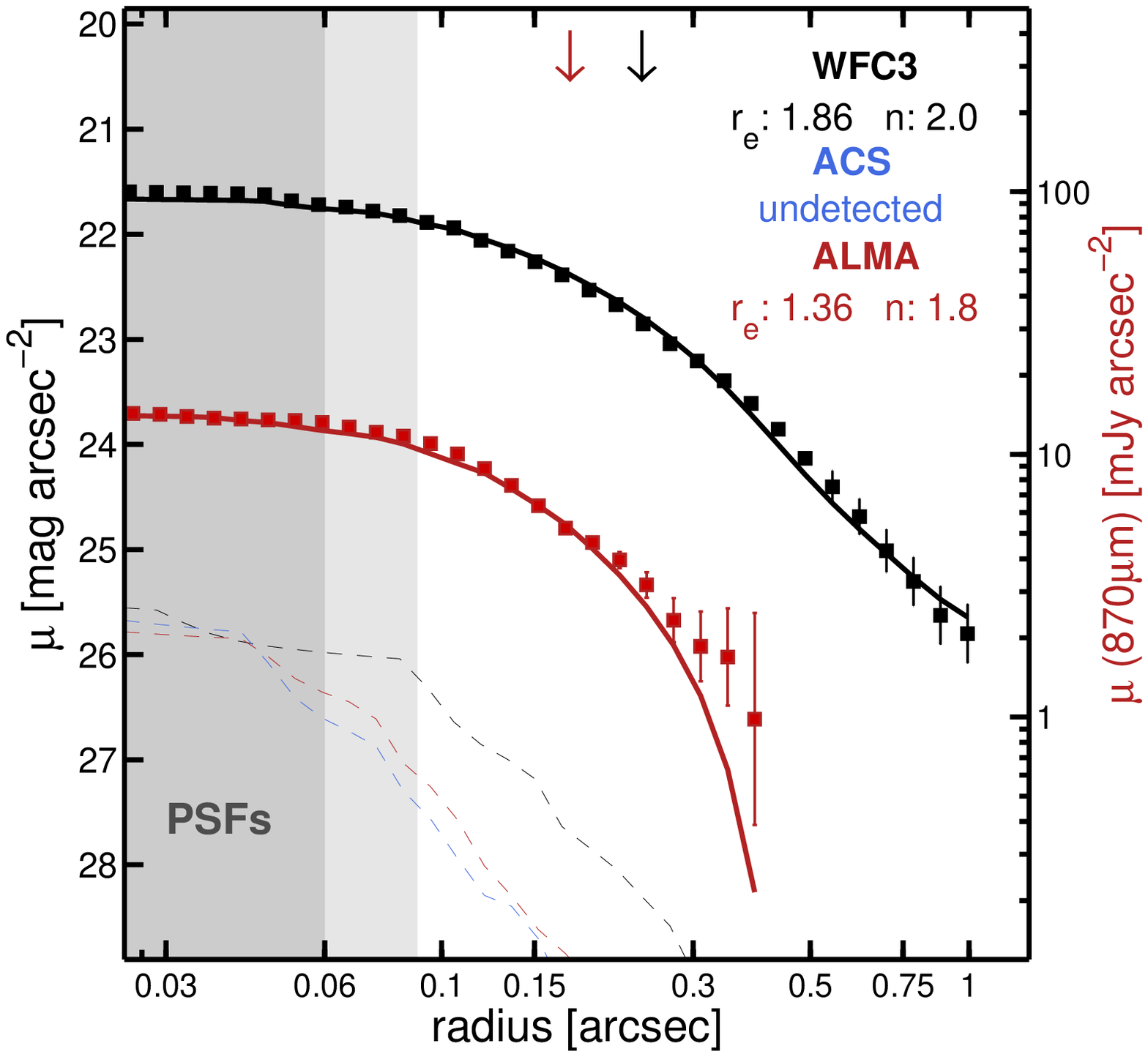}
\caption{\label{profiles} $2\farcs5\times2\farcs5$ images and surface
  brightness profiles (un-corrected for the PSF) of the compact SFGs
  in ACS/F850LP, WFC3/F160W and ALMA~870~$\mu$m (with the F160W
  contours shown in black). The surface brightness profiles (squares)
  are measured along concentric ellipses which follow the geometry of
  the best-fit S\'ersic model (solid lines). The ALMA~870~$\mu$m
  profiles are scaled down arbitrarily with respect to the HST data
  (see right y-axis). The ALMA and ACS images have similar spatial
  resolution (FHWM~$\sim0\farcs12$) and are slightly smaller than for
  WFC3 (FHWM~$\sim0\farcs18$). The dashed lines show the PSF profiles
  and the shaded regions show the extent of their HWHM in F850LP and
  ALMA (dark grey) and F160W (light grey). The deconvolved GALFIT
  effective radii in kpc are indicated with arrows. The profiles are
  shown up to the radius where the errors become significant,
  typically $\sim1''$ in F160W and $\sim0\farcs4$ in ALMA.}
\end{figure*}

The sub-mm observations of the 6 targets were taken as part of an ALMA
cycle-2 campaign (ID: 2013.1.00576.S; PI: G. Barro) aimed at studying
the dust emission continuum in compact SFGs at $z=2-3$. The
observations were carried out on 2015-08-29 and 2015-09-07 in band 7
using four spectral windows in the largest bandwidth mode.  The
on-source integration time was 1800s in the longest array
configuration, C34-7. Flux, phase, and band-pass calibrators were also
obtained, for a total time of $\sim3$~hr. We used the CASA software to
process and clean the data. The cleaning algorithm was run using a
natural weighting for the {\tt u-v} visibility plane. The average
angular resolution of the observations is
FWHM~$=0\farcs14\times0\farcs11$, with a major-axis position angle
ranging from $3^{\circ}-65^{\circ}$ (see Figure~\ref{profiles}). The
rms noise of the observations is $\sigma=40$~$\mu$Jy/beam or
$2.4$mJy/arcsec$^{2}$. The depth of the ALMA observations allow
reliable measurements of the surface brightness profile at a 3$\sigma$
level down to a radius of at least 5$\times$ the half-width at
half-maximum of the ALMA clean beam. We find and correct an average
systematic offset between the HST and ALMA astrometry\footnote{The
  offsets are consistent with recent results from a JVLA survey of
  GOODS-S (Rujopakarn in prep.)} of $\Delta$RA~=$-0\farcs08$ and
$\Delta$DEC~$=0\farcs$27 with an rms$\sim0\farcs06$.

\section{Optical/NIR and MIR/submm SED fits: M$_{\star}$, SFR, M$_{\rm dust}$ and $M_{\rm gas}$}
\subsection{models and assumptions}

We fit the optical/NIR SEDs to calculate stellar masses using FAST
\citep{fast} and assuming \cite{bc03} stellar population synthesis
models, and the \cite{calzetti} dust law with attenuation
$0<A_{V}<4$. We also assume an exponentially-declining star formation
history with timescale $\tau$ and age $t$ (see \citet{santini15} for
more details).

We fit the mid-to-FIR SEDs to the dust emission templates of
\citet{ce01}, \citet{dh02}, and \citet{rieke09}. Moreover, we fit the
models by \citet[][DL07]{dl07} to estimate the physical properties of
the dust.  In these models, dust is exposed to a range of starlight
intensities indicated by a scale factor $U$. The majority of the dust
is heated by a constant intensity $U_{\rm min}$, while a smaller
fraction $\gamma$ of the dust is exposed to variable intensities
ranging from $U_{\rm min}$ to $U_{\rm max}$. The parameter, $q$,
controls the fraction of the dust grains in the form of polycyclic
aromatic hydrocarbon grains (PAH).  We find the best-fit DL07 models
and the corresponding confidence intervals by exploring the parameter
space using the Python Markov-Chain Monte Carlo (MCMC) package {\tt
  emcee} \citep{mcmc}.

We also estimate the molecular gas content using the gas-to-dust ratio
by assuming $\delta_{\rm GDR}$\mdust$= M_{\rm gas}$. The value of
$\delta_{\rm GDR}$ depends primarily on the metallicity of the galaxy
(e.g., \citealt{sandstrom13}). We estimate $\delta_{\rm GDR}(Z)$ using
the mass-metallicity (MZR) relation at $z\sim2$ of \citet{steidel14},
and the empirical calibration of \citet{magdis12}. For the relatively
small range of galaxy masses in our sample the average metallicity is
Z~=~12+log(O/H)~$=8.57$ which implies gas-to-dust ratios of
$\delta_{\rm GDR}\sim100$.

We compute the total SFR by adding the unobscured and obscured star
formation, traced by the UV and IR emission, respectively, following
\citet[][see also \citealt{bell05}]{ken98}.
\begin{equation}\label{SFRform}
SFR_{\mathrm{UV+IR}}=1.09\times10^{-10}(L_{\mathrm{IR}}+3.3L_{2800})[M_{\odot}/\rm yr]
\end{equation}

\noindent
where $L_{\mathrm{IR}}$ is the total IR luminosity ($L_{\rm{IR}}\equiv
L(8-1000~\mu$m)) derived from the average value of the best-fit
templates to the 4 dust emission libraries, and $L_{2800}=\nu
L_{\nu}(2800)$ is estimated from the best-fit SED models.

\subsection{Integrated IR-, UV- SFRs, dust and gas masses}

The compact SFGs exhibit high IR luminosities ranging from L$_{\rm
  IR}=10^{12.03-12.80}$~L$_{\odot}$ and SFR~$=150-730$~\suny~that
straddle the SFR main sequence at \lmass~$\sim$11
(Figure~\ref{selectiondiag}). Two of the galaxies have slightly larger
SFRs than the median of SFR~$\sim$300~\suny, yet still within
$5\times$ of the SF-MS. The total SFR is strongly dominated by the IR
emission. The average ratio of integrated SFR$_{\rm IR}$/SFR$_{\rm
  UV}$~$=70-100$ implies optical attenuations of $A_{\rm
  V}\gtrsim2$~mag. However, the values determined from SED fitting are
only $A_{\rm V}\sim1.3-1.6$~mag.

The IR-SFRs determined from the different dust template libraries are
in excellent agreement, with a median difference and 1$\sigma$ scatter
of $\Delta$SFR~$=0.01\pm0.13$~dex.  The SFRs are also consistent with
estimates based on the MIPS 24~$\mu$m flux alone, using the empirical
relation from \citet{wuyts11a}, $\Delta$SFR~$=0.02\pm0.26$~dex. The
two X-ray detected AGNs in the sample (GDS-9834 and GDS-11701) exhibit
slightly higher MIPS 24~$\mu$m SFRs, likely as a result of hot dust
emission from the AGN at shorter wavelengths (see also Barro et
al. 2014a). We find similar SFR values when the 24~$\mu$m flux is
excluded from the fits.

The dust masses range from $M_{\rm dust}=10^{8.05-9.29}$~M$_{\odot}$,
and are larger for the 2 galaxies with the highest SFRs. The gas
masses determined from $\delta_{\rm GDR}$ and the MZR relation
indicate gas fractions of the order of f$_{\rm gas}=M_{\rm
  gas}/(M_{\rm gas}+M_{\star})=0.47^{+0.19}_{-0.15}$, consistent with
previous works (\citealt{tacconi10}).  The average depletion time,
assuming no further gas replenishment, is relatively short, $t_{\rm
  dpl}=M_{\rm gas}/{\rm SFR}=230^{+90}_{-120}$~Myr.

\begin{figure*}[t]
  \centering
  \includegraphics[width=7.6cm,angle=0.]{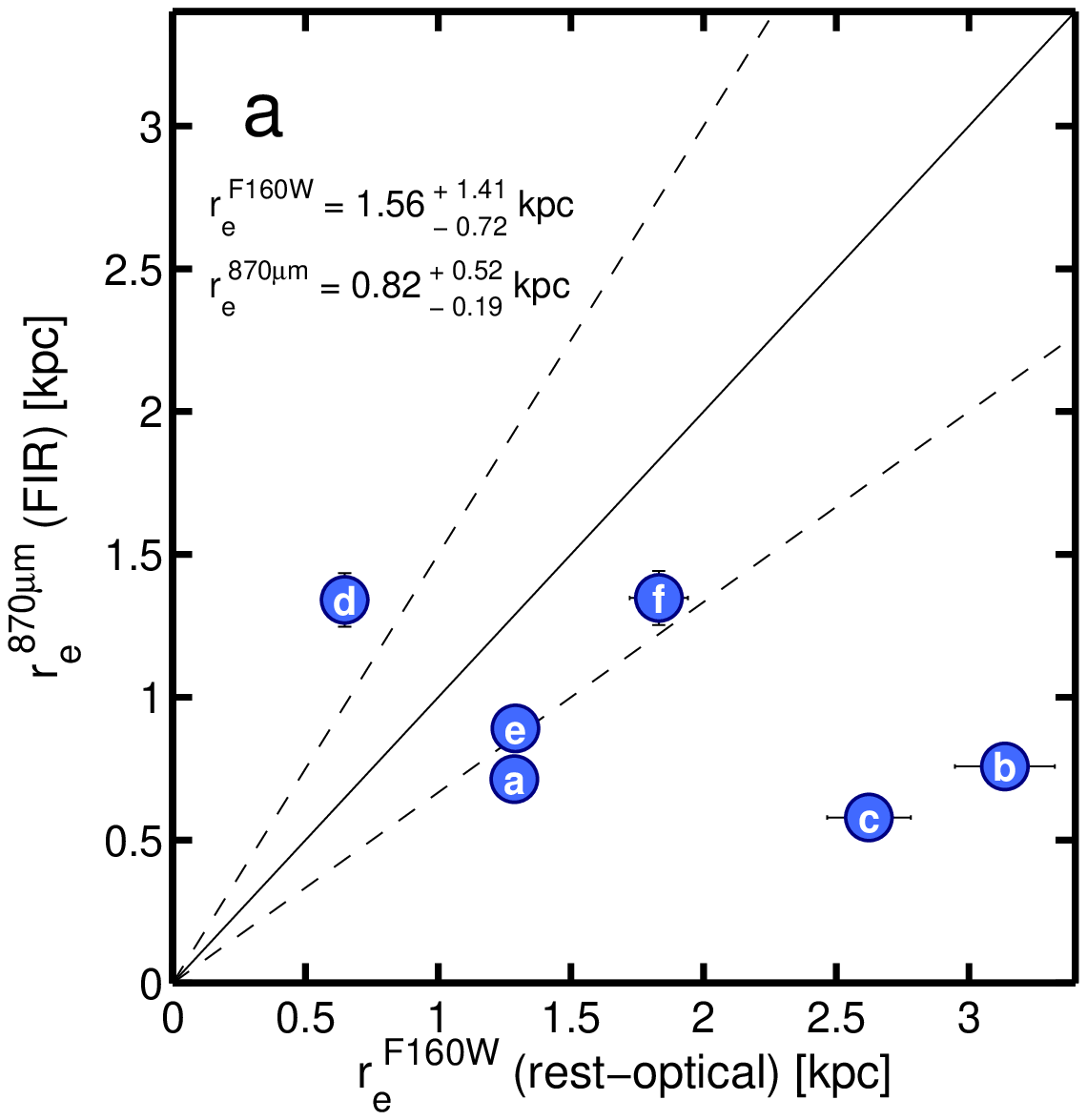}
  \hspace{1.3cm}
  \includegraphics[width=8.0cm,angle=0.]{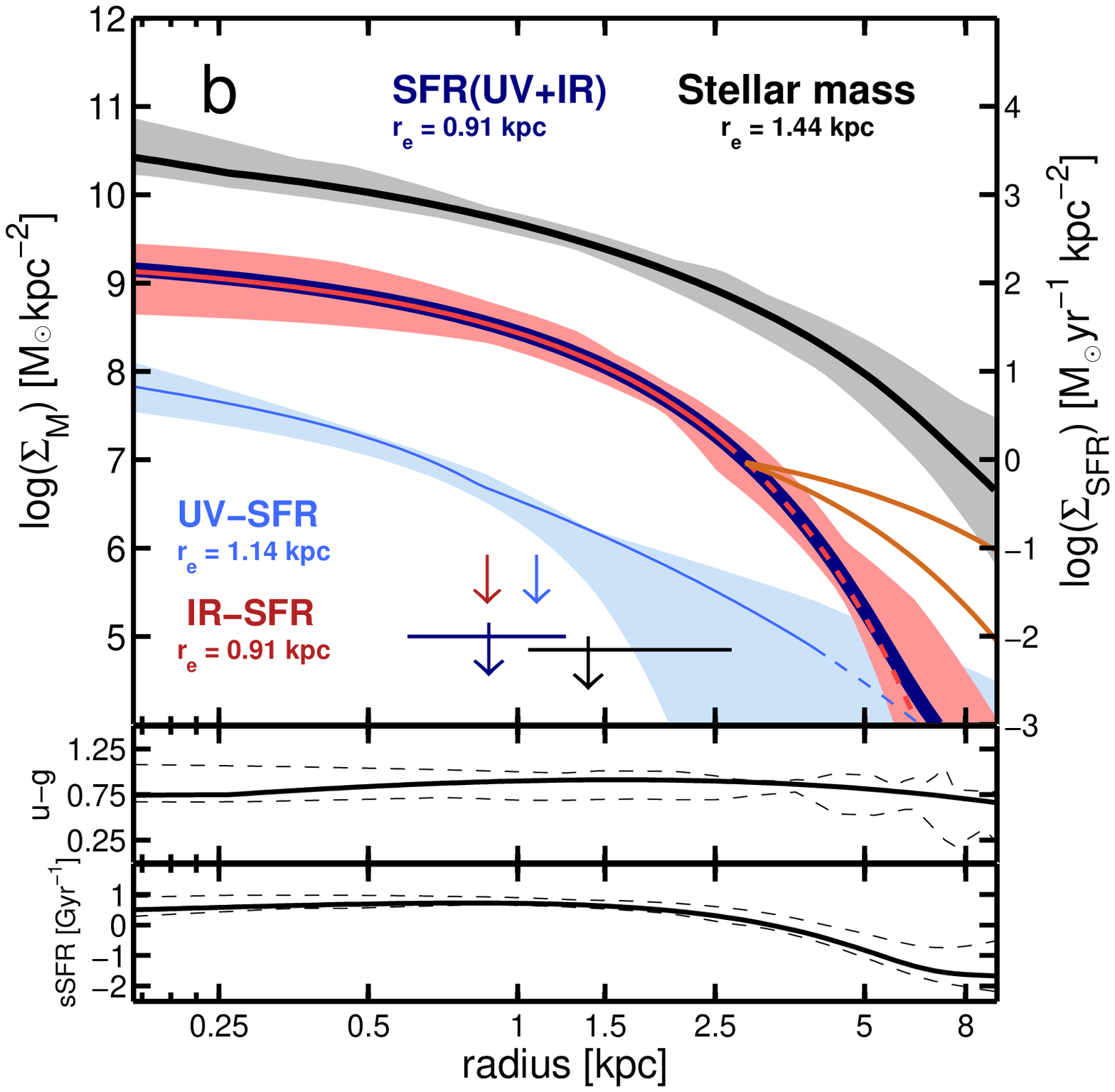}
  \caption{\label{stack} {\it Left:} Comparison of the rest-frame
    optical (WFC3/F160W) and FIR (ALMA~870~$\mu$m) effective radii of
    compact SFGs. The solid line indicates the 1:1 relation, the
    dashed lines show the 1.5$\times$ size ratios. The FIR sizes are
    $\sim$1.6$\times$ smaller than the optical sizes and they exhibit
    a tighter distribution around $r_{e}\sim1$~kpc. {\it Right:} Mean
    de-convolved stellar mass (black line) and SFR (blue and red
    lines) density profiles of compact SFGs. The shaded regions
    indicate the 1$\sigma$ dispersion. The dashed lines indicate
      the S\'ersic fit to SFR profile below the UV and FIR detection
      limits. The orange lines show possible IR-SFR profiles
      undetected by ALMA. The arrows show the mean effective radii
    and the horizontal bars indicate their lower/upper limits
    determined from the $\pm1\sigma$ profiles. The bottom panels show
    the $u-g$ color profile and the sSFR profile.}
\end{figure*}

\section{Structural properties}\label{sectsizes}
\subsection{UV, Optical and FIR surface brightness profiles}

Figure~\ref{profiles} shows the images and surface brightness profiles
of the galaxies in WFC3/F160W, ACS/F850LP and ALMA~870~$\mu$m. At
$z\sim2.5$, these bands probe the rest-frame UV, optical, and FIR,
respectively. We account for the different spatial resolution of each
dataset by modeling the shape of the two-dimensional surface
brightness profiles using GALFIT \citep{galfit}. The half-light radii
and S\'ersic indices, $n$, are determined using a single component
fit. For the HST images, the PSFs are created with TinyTim
\citep{tinytim} as detailed in \citet{vdw12}. For ALMA, we used the
synthetic PSF generated by CASA. The depth of the ALMA observations
($t_{\rm exp}\sim$~30~min) provides a smooth, uniform coverage of the
PSF.  Figure~\ref{profiles} indicates the best-fit $r_{\rm e}$
(arrows) and $n$ in each band.

The mean radius containing 95\% of the light from the observed FIR
profiles is $\sim4\times$ smaller than that of the optical
profile. The deconvolved, FIR profiles are also more compact than the
optical in 5/6 galaxies. The exception is GDS-3280, which has the
smallest optical size and a high $n$, while the morphology in ALMA
exhibits some slight asymmetries. The ratio of the mean effective
radii is $\langle r_{\rm e,F160W}\rangle/\langle r_{\rm e, 870\mu
  m}\rangle=1.9$. Despite being more compact, the FIR profiles have,
on average, lower (disk-like) S\'ersic indices with $n_{870\mu
  m}\sim1$ than the optical profiles with $n_{\rm
  F160W}\sim2$. Figure~4a compares the FIR and optical $r_{\rm e}$ ans
shows no clear correlation. Furthermore, the FIR sizes exhibit a
tighter size distribution, that suggests a relatively homogeneous
population of compact, dusty, starbursts confined to the nuclear
regions. The small star-forming regions with integrated SFRs
  consistent with the SF-MS imply that the nuclear starbursts have
  $\langle \Sigma_{\rm SFR}\rangle= SFR / \pi r_{\rm e}^{2}$ up to
  $25\times$ larger than a typical SF-MS disk with $r_{\rm e, SFR}
  \sim 5$~kpc \citep{nelson15}.

The remarkable compactness of the dust continuum emission is
consistent with recent results on SMGs and other IR-bright galaxies
which report small FIR sizes (Gaussian FHWM~$\sim0\farcs12$) compared
to the typical rest-frame optical sizes of SFGs at $z\sim2-3$
(\citealt{simpson15}; \citealt{ikarashi15};
\citealt{tadaki15}). Nonetheless, joint studies of both the UV- and
IR- SFRs, and the stellar mass profiles are required to fully
constrain the regions where stars are being formed and to understand
their role in the structural evolution of SFGs.

\subsection{Stellar mass, UV- and IR- SFR surface density profiles}

Figure~4b shows the average stellar mass and SFR profiles of the
compact SFGs as computed from their deconvolved, surface density
profiles. The UV-SFR and IR-SFR profiles are determined by scaling the
rest-frame luminosity profiles, probed by ACS/F850LP and
ALMA~870~$\mu$m, to an integrated SFR using the conversion in
Equation~\ref{SFRform}. The stellar mass profiles are determined from
the rest-frame optical luminosity probed by WFC3/F160W (approximately
g-band, $L_{\rm g}$) by using an empirical correlation between the
stellar mass-to-light ratio, $M_{\star}/L_{\rm g}$, and the rest-frame
$(u-g)$ color as determined from F125W and F160W (see e.g.,
\citealt{szo12}). We account for resolution effects in the color
profile by using the best-fit GALFIT models of the F125W and F160W
brightness profiles.

The average (UV+IR) SFR profile is approximately $1.5\times$ more
concentrated than the average stellar mass profile and it is strongly
dominated by the IR emission (UV/IR~$\gtrsim100$) up to
$r\sim5$~kpc. The specific SFR (sSFR~$=$~SFR/M) is highest at
$r\lesssim2.5$~kpc and thus imply that most of the stellar mass growth
is taking place within the inner few kpc of the galaxy. At
$r\gtrsim5$~kpc, the sSFR is $\sim$100$\times$ lower thus indicating
that the SFR has an almost negligible contribution to the stellar mass
growth at large radii.

Note that the UV- and IR- SFR profiles are detected only up to
$r\sim3-4$~kpc ($\Sigma_{\rm SFR}\sim1 M_{\odot}$~kpc$^{-2}$ and
$\sim0.1 M_{\odot}$~kpc$^{-2}$). Therefore, the results at larger
radii are based on the best-fit S\'ersic profiles. Nonetheless, we
verify that even if the IR-SFR profiles had secondary components
undetected by ALMA with $\Sigma_{\rm SFR}(r = 8~\rm
kpc)\sim$10$\times$ and 100$\times$ lower than the detection limit
(orange lines in Figure~\ref{stack}b), the $r_{\rm e}$ would only
increase by $\sim$5\% and 20\% and thus the SFR would still be more
concentrated than the stellar mass.

\section{Discussion}

\begin{figure*}
  \centering
  \includegraphics[width=6.15cm,angle=0.]{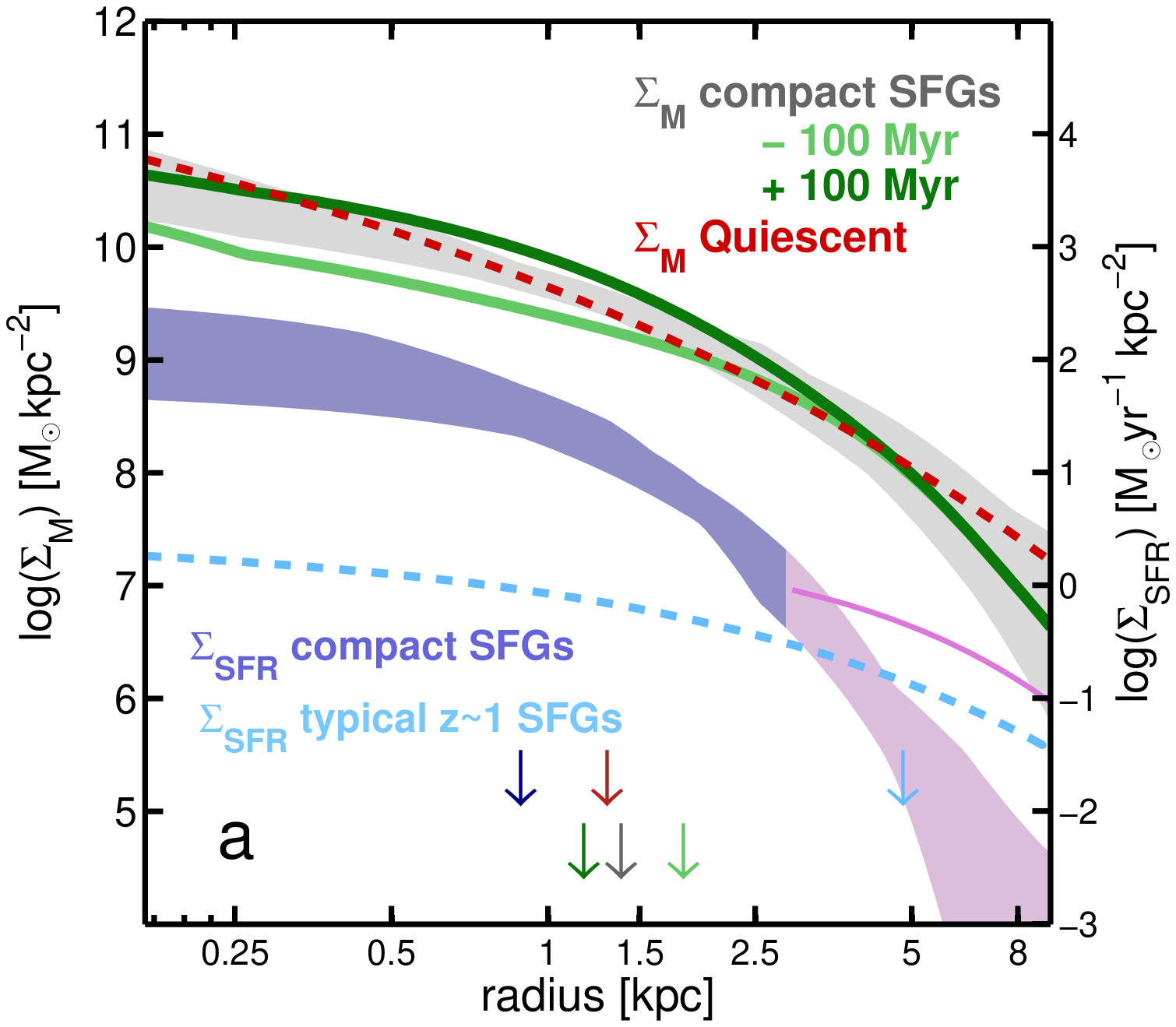}
  \hspace{0.3cm}
  \includegraphics[width=5.5cm,angle=0.]{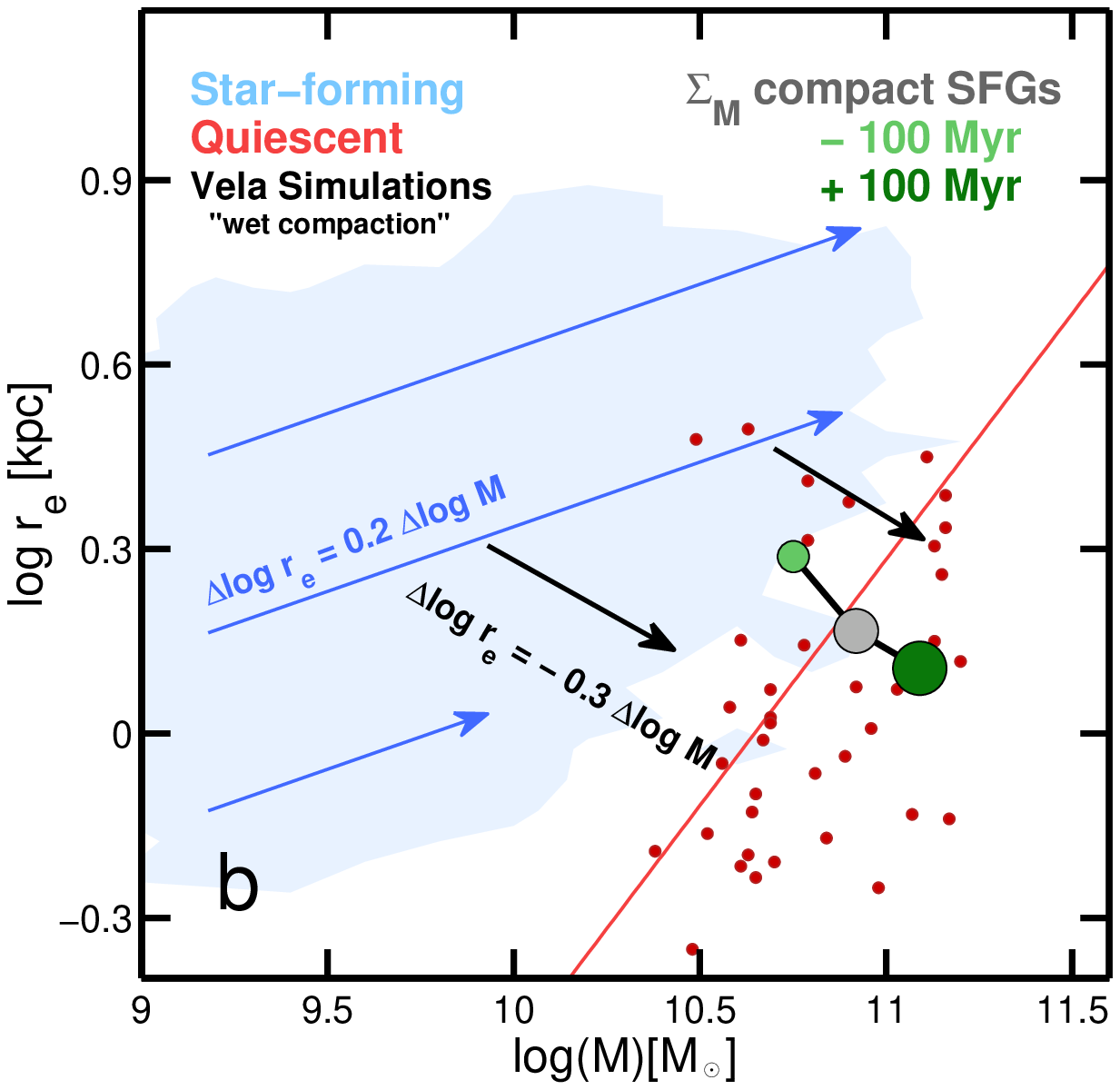}
  \includegraphics[width=5.5cm,angle=0.]{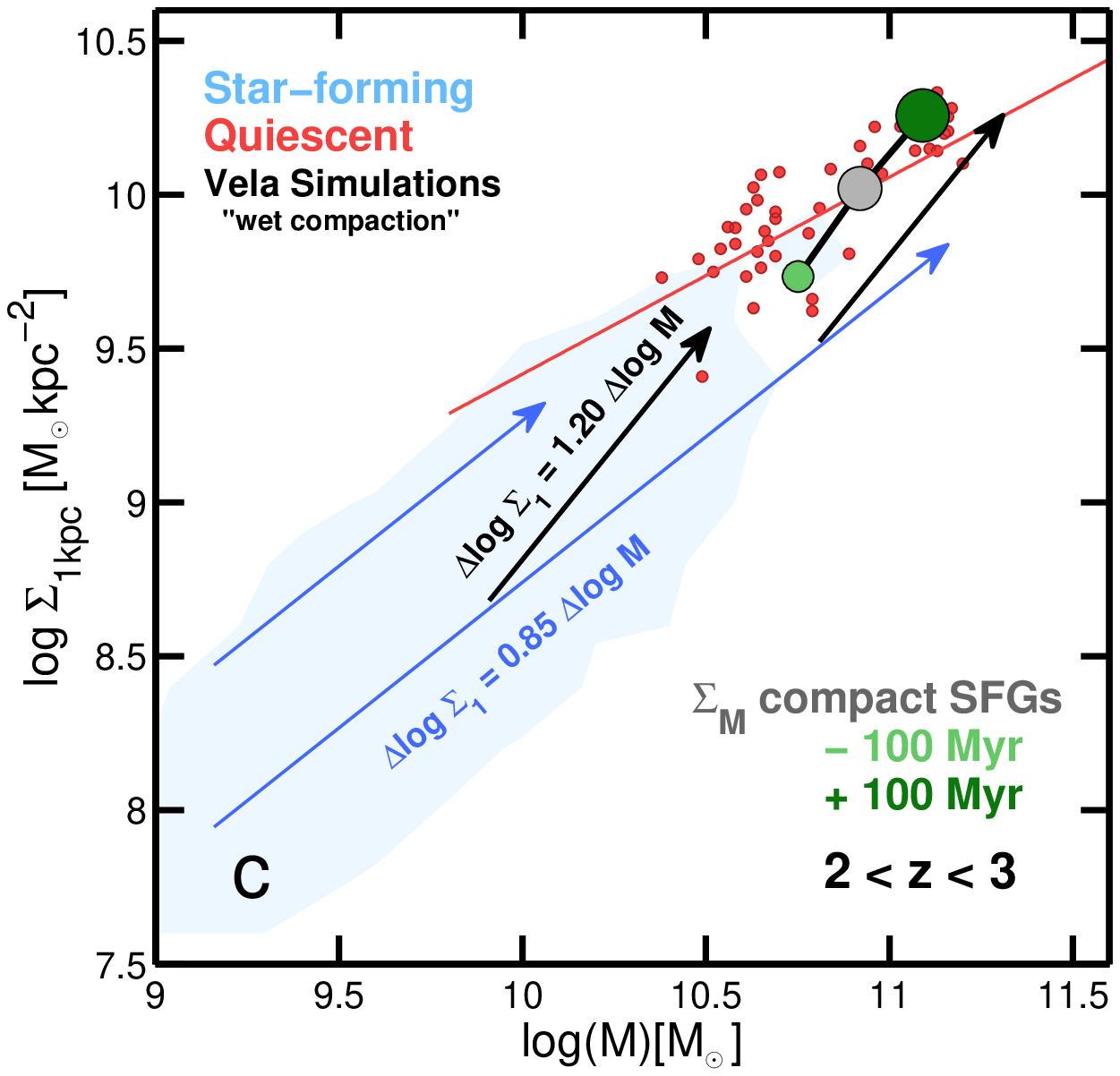}
  \caption{\label{evoltracks} {\it Left:} Evolution of the
    $\Sigma_{\rm M}$ profile (grey contour) of compact SFGs assuming
    that their $\Sigma_{\rm SFR}$ (blue contour) remains constant
    during $\Delta t=\pm100$~Myr (light-to-dark green lines). The
    magenta contour indicates the region below the ALMA detection
    limit. The magenta line shows a possible extended star-forming
    component (see \S~4.2). The red dashed line shows the mean stellar
    mass profile of compact quiescent galaxies at $z\sim2$. The cyan
    dashed line indicates the $\Sigma_{\rm SFR}$ of typical SFGs at
    $z\sim1$ from \citet{nelson15} (see also \citet{tacchella15} at
    $z\sim2$). The arrows indicate the $r_{\rm e}$. {\it Middle:} The
    blue contour and the red points indicate the loci of star-forming
    and quiescent galaxies at $z\sim2$, and the blue and red lines
    depict their best-fit scaling relations from \citet{vdw14}. The
    light-to-dark green circles indicate the size-mass evolution of
    the stellar mass profile in panel a. The black arrows show the
    direction of the structural evolution in the Vela simulations
    during the ``wet compaction'' phase. {\it Right:} Same as middle
    panel but showing the \sigone-mass evolution. The average
    relations for SFGs and quiescent galaxies at $z\sim2$ are from
    Barro et al. (2016b).}
\end{figure*}

In a simplified picture of galaxy growth, the average structural
evolution of SFGs proceeds roughly along their well-defined scaling
relations (blue arrows in Figure~\ref{evoltracks}; e.g.,
\citealt{dokkum15}; Barro et al. 2016b). In this picture, massive
compact quiescent galaxies at $z\sim2$ would be descendants of smaller
SFGs at higher-z that achieve such high stellar densities by
continuously growing in stellar mass and size fueled by extended SFR
profiles. Alternatively, these SFGs could deviate from the smooth
track due to dissipative processes that would rapidly increase their
concentrations and potentially decrease their half-mass radii in
strong nuclear starbursts (\citealt{dekel13b};
\citealt{wellons15}). The {\it secular} and dissipation-driven
scenarios are not mutually exclusive. However, we aim to understand
whether the massive dense cores of compact quiescent galaxies are
primarily formed in dissipative processes.

The strong nuclear starbursts embedded in larger stellar mass profiles
found in compact SFGs are indeed an excellent match to the
dissipation-driven scenario. The light-to-dark green lines and circles
in Figure~\ref{evoltracks} show the predicted change in the stellar
mass profile and the evolutionary tracks in $r_{\rm e}$ and central
mass density for compact SFGs due to star formation, assuming that
their SFR profiles remain constant during $\Delta t=200$~Myr
(approximately $t_{\rm dpl}$). The significant stellar mass growth
within the inner $r\lesssim2$~kpc decreases the half-mass radius by
$1.6\times$ from $r_{\rm e,mass}=1.9$ to $1.2$~kpc, while the central
density within $r\leq1$~kpc increases by $\sim4\times$ from
log(\sigone)~$=$~9.7 to 10.3~M$_{\odot}$~kpc$^{-2}$. If compact SFGs
had more extended star formation at $r\gtrsim3$~kpc the evolution of
\sigone~would be the same, while $r_{\rm e, mass}$ would decrease less
(e.g., $\sim$7\% for the magenta line).

These evolutionary tracks are very similar to the predictions of the
Vela simulations during the ``wet compaction'' phase (black arrows;
e.g., \citealt{zolotov15}; \citealt{tacchella16}) and contrast with
the expected evolution for typical SF-MS galaxies, which have extended
SFR profiles with $\sim100\times$ lower central $\Sigma_{\rm SFR}$
(cyan line in Figure~\ref{evoltracks}a) and thus favor a more gradual
increase of \sigone~and a positive size evolution.

The short depletion times of compact SFGs and the similarity with the
mean stellar mass profile of quiescent galaxies at $z\sim2$ (red
dashed line in Figure~5a) suggest that the nuclear starburst is
unlikely to continue for more than a few hundred Myrs, either because
no further gas is accreted into the galaxy center or because the dense
stellar component stabilizes the gas to prevent further star formation
and eventually leads to galaxy quenching. This scenario is consistent
with previous results indicating that the formation of a dense core
precedes the shut down of star formation (e.g., \citealt{cheung12};
\citealt{dokkum14}), and suggests that, at high redshift, both
quenching and the dense cores are simultaneous consequences of
enhanced periods of nuclear star formation that cause a rapid
depletion of the gas reservoirs.

\section*{Acknowledgments}

GB and MK acknowledges support from HST-AR-12847 and the Hellman
Fellows Fund. PGP-G acknowledge ssupport from grant AYA2015-70815-ERC.

\bibliographystyle{aa}

\clearpage
\label{lastpage}

\end{document}